\documentclass[apj]{emulateapj} 
\usepackage{amsmath}
\usepackage{graphicx}
\usepackage{natbib}
\usepackage{hyperref}

\shorttitle{Pseudo-Newtonian equations in stationary space-times}
\shortauthors{Witzany \& L{\"a}mmerzahl}

\renewcommand{\d}{\mathrm{d}}

\begin{document}
\title{Pseudo-Newtonian equations for evolution of particles and fluids in stationary space-times}


\author{Vojt{\v e}ch Witzany}
\email{vojtech.witzany@zarm.uni-bremen.de}
\affiliation{ZARM, Universit{\"a}t Bremen, Am Fallturm, 28359 Bremen, Germany}
\affiliation{Institute of Theoretical Physics, Faculty of Mathematics and Physics, Charles University in Prague, Prague, Czech Republic}
\author{Claus L{\"a}mmerzahl}
\email{claus.laemmerzahl@zarm.uni-bremen.de}
\affiliation{ZARM, Universit{\"a}t Bremen, Am Fallturm, 28359 Bremen, Germany}
\affiliation{Institut f{\"u}r Physik, Universit{\"a}t Oldenburg, 26111 Oldenburg, Germany}


\begin{abstract}

Pseudo-Newtonian potentials are a tool often used in theoretical astrophysics to capture some key features of a black-hole space-time in a Newtonian framework. As a result, one can use Newtonian numerical codes, and Newtonian formalism in general, in an effective description of important astrophysical processes such as accretion onto black holes.

In this paper we develop a general pseudo-Newtonian formalism which pertains to the motion of particles, light, and fluids in stationary space-times. In return, we are able to assess the applicability of the pseudo-Newtonian scheme. The simplest and most elegant formulas are obtained in space-times without gravitomagnetic effects, such as the Schwarzschild rather than the Kerr space-time; the quantitative errors are smallest for motion with low binding energy. Included is a ready-to-use set of fluid equations in Schwarzschild space-time in Cartesian and radial coordinates.

\end{abstract}




\section{Introduction}
\label{sec:intro}


Until now, all gravity related observations can be fully described within General Relativity (GR). There is no single gravitational phenomenon which is in contradiction to GR. Within GR the gravitational field is given by the Einstein field equations which are highly complicated and can be solved exactly only for a few highly symmetric configurations. Even for the two body system no exact solution can be found, contrary to the situation in Newtonian gravity. In order to cope with such more complicated situations, various analytic approximation schemes have been developed. 

The earliest and most prominent approximation scheme, used already by \citet{einstein} for the first computation of the perihelion shift of Mercury, is the post-Newtonian approach (see e.g. \citet{blanchetlivrev}). This approach is essentially characterized as a formal expansion of the field equations and the equations of motion in terms of orders of the ``slowness'' of the system $v/c$, where $v$ is some characteristic velocity of the system. A second prominent approximation scheme is the post-Minkowski approach which is a weak field approximation expanding in terms of deviations from the flat space-time background and which is applicable to any velocity of the constituents of the systems. 

However, for bound motion, such as is the case of binary stars, the deviation from the flat background and the typical velocity of the objects are intimately related. In the case of a test body in a circular orbit of radial Schwarzschild coordinate $r_\mathrm{c}$ around a Schwarzschild black hole, the test body's velocity is $v/c = \sqrt{GM/(c^2 r_\mathrm{c})}$ and the deviation from the flat background at $r_\mathrm{c}$ can be expanded in terms of $GM/(c^2 r_\mathrm{c})$. Hence, a post-Minkowski expansion naturally leads to a post-Newtonian one and vice versa (see e.g. \citet{sasakilivrev}). Thus, both the post-Newtonian and the post-Minkowski expansion schemes applied to bound motion are {\em weak-field} approximations and are not suited to describe bound motion in the very vicinity of a black hole, or more generally, in the strong gravity regime.

Nevertheless, astrophysicists often have to describe, e.g., an accretion disc extending in terms of its inner radius up to the innermost stable circular orbit (ISCO) $r_\mathrm{ISCO} = 6GM/c^2$of the black hole, or even the photon sphere $r_\mathrm{ps}= 3 GM/c^2$ (see e.g. \citet{abramowiczlivrev}). It is obvious that only an extremely careful and laborious post-Newtonian or post-Minkowski expansion would provide a satisfactory description of the motion in the very vicinity of the black hole.
On the other hand, the {\em outer} radius of the accretion disc often extends up to hundreds of $GM/c^2$ where  relativistic effects become completely negligible. 

In other words, the largest part of the accretion process is governed by Newtonian physics and only in the very last few percents of accretion the behavior of the accreted matter as a test field on a strongly curved general-relativistic background becomes important. However, these last stages are essential for the global structure of a steady accretion disc, because of the instability of the disc beyond the ISCO and the precise energetics near the black hole determining the amount of energy radiated away during the accretion process. 

Hence, to model accretion on a black hole, we would ideally like a dynamical description which is mostly Newtonian but reproduces some of the characteristic features of motion near the black hole. Precisely this kind of model was given first by \citet{paczynsky} by placing the fully Newtonian accretion disc into a non-physical gravitational field with the potential
\begin{equation}
\Phi_\mathrm{PW} = - \frac{GM}{r- 2GM/c^2}\,.
\end{equation}
Obviously, this potential very quickly obtains the Newtonian asymptotics $\sim -GM/r$ as $r\gg GM/c^2$ but a quick computation of its Laplacian shows that it would have to be generated by infinite densities of negative matter. On the other hand, it has an ISCO at $r=6 GM/c^2$ with specific binding energy $\tilde{E}=1-\sqrt{8/9}$. Since these are the same values of the coordinate radius and of the binding energy of the ISCO as in the Schwarzschild space-time, this so-called Paczy{\`n}ski-Wiita potential can be used as an effective model of the static black-hole field (see \citet{abramowiczPW} for a review of other properties). 

Since the publication of the Paczy{\'n}sky-Wiita potential, the approach of reproducing some of the characteristic features of a selected class of orbits within an ad-hoc Newtonian framework has been called the ``pseudo-Newtonian" (pN) approach (see introduction of \citet{tejeda1} or \citet{artemova} for a review). 

Even though pseudo-Newtonian potentials have been proposed for over 35 years \citep{paczynsky, nowak,artemova,semerak1999,mukhopadhyay2002,mukhopadhyay2003,chakrabarti2006,ghosh2007,wegg,tejeda1,sarkar,tejeda2,ghosh,paperIV}, until recently the potentials were not able to accurately reproduce properties of general orbits or to accurately describe the field of a rapidly spinning black hole. However, \citet{tejeda1,tejeda2} proposed a class of generalized (velocity-dependent) pseudo-Newtonian potentials accurately describing the motion of quite general test-particles in the Schwarzschild and generally any spherically symmetric space-time (the same result on spherically symmetric space-times was almost simultaneously given by \citet{sarkar}). The pattern in the formulation is remarkably simple, the pseudo-Newtonian Lagrangians of  \citet{tejeda1,tejeda2,sarkar} can be all given within a single formula
\begin{equation}
L = \frac{1}{2} \left( \frac{\dot{r}^2}{f(r)^2} + \frac{\dot{\vartheta}^2 + \sin^2 \! \vartheta \, \dot{\varphi}^2}{f(r)} \right) + \frac{1}{2} f(r)\,, \label{eq:spherprev}
\end{equation}
where $f(r)=-g_{tt}=1/g_{rr}$ and $t,\,r,\,\varphi,\,\vartheta$ are Schwarzschild-like spherical coordinates. The key step in the derivation of these pseudo-Newtonian Lagrangians always seems to be the assumption that the specific energy of the test particle $-u_t$ is approximately 1. Additionally, all the derivations have been done for massive particles. As a result, these Lagrangians give exactly the position of the ISCO, the marginally bound ($-u_t=1$) circular orbit, the formal angular momentum distribution over circular orbits and a general quantitative agreement with the exact relativistic case.

However, this pattern does not include spinning black holes, i.e. the Kerr metric. For the Kerr space-time, a number of proposals exist, usually fitting the potential by the behavior of some set of orbits or by ``reading off" a potential from the equations of motion or the metric \citep{semerak1999,mukhopadhyay2002,mukhopadhyay2003,chakrabarti2006,ghosh2007}. 

The only proposal for a pseudo-Newtonian description of a spinning black holes which somehow follows the line of reasoning of \citet{tejeda1} is that of \citet{ghosh}, where the authors derived a generalized pseudo-Newtonian potential for test-particles in the equatorial plane of a slowly spinning Kerr black hole by utilizing a ``low-energy limit" in one of the steps of the derivation. The restriction of the particles in the equatorial plane seems to play mainly the role of convenience during the derivation but the limit on spin of the black hole is set because of emergent singular behavior of important circular orbits, namely the marginally bound and marginally stable orbit (see Subsection \ref{subsec:ghoshremarks} for more details). 

Even though the development of pseudo-Newtonian potentials is oriented towards magneto-hydrodynamics of a plasma near a black hole, none of the papers  \citep{tejeda1,tejeda2,sarkar, ghosh} have discussed the applicability of their framework in this context.

To conclude our discussion, we give a set of questions which have not been so far addressed in the literature: Are the new velocity-dependent potentials also applicable for null geodesics, i.e. computations of gravitational lensing or black-hole shadows? How does this formalism implement additional forces such as electromagnetism? Is it correct to use these Newtonian-like Lagrangians (\ref{eq:spherprev}) along with non-modified Newtonian fluid dynamics as was done e.g. by \citet{bonnerot}? Is there a deeper pattern in the way the Lagrangians are formulated and can we perhaps extrapolate it to highly spinning black holes and off-equatorial particles near them? This paper partially resolves these questions.

In Section \ref{sec:pNHam} we derive a general pseudo-Newtonian Hamiltonian valid in any stationary space-time and specify the exact relationship the corresponding trajectories have to relativistic geodesics in the original space-time. In Section \ref{sec:timediag} we focus on its properties in the most elegant and simple case of static space-times, and in Section \ref{sec:spher} we show that the herein presented Hamiltonian encompasses all the recently published velocity-dependent pseudo-Newtonian potentials (\ref{eq:spherprev}). 

As far as concerns new applications of this work, Section \ref{sec:applications} gives also a pseudo-Newtonian Hamiltonian for charged particles and derives pseudo-Newtonian fluid equations in static space-times. The last Section \ref{sec:kerr} then discusses the properties of the pseudo-Newtonian Hamiltonian for spinning black holes.


\section{Pseudo-Newtonian Hamiltonian}
\label{sec:pNHam}


We use the $G=c=1$ geometrized units and the $-+++$ signature of the metric. Space-time coordinates are labeled by Greek letters, spatial coordinates by roman letters. At certain instants we will switch to SI units and indicate so.
\subsection{Flat space-time}
Consider the Lagrangian and respective Hamiltonian of the motion of a test particle in flat space-time in Cartesian coordinates
\begin{align}
L=\frac{1}{2} \eta_{\mu \nu} u^{\mu} u^\nu\,,\\
H=\frac{1}{2} \eta^{\mu \nu} u_{\mu} u_\nu\,,
\end{align}
where $u^\mu\equiv \d x^\mu /\d \tau$ is the four-velocity of the particle and $u_\mu$ is canonically conjugate to $x^\mu$. This description gives the trajectory as parametrized by proper time $\tau$ rather than the time in the laboratory frame $t=x^0$. However, since both the Lagrangian and Hamiltonian have the same on-shell value $H=L=-1/2$ for any massive particle, we can obtain a Hamiltonian for trajectories parametrized by time in the laboratory frame instead of proper time
\begin{align}
H_t=-u_t=\sqrt{1 +\sum_i (u_i)^2}\,,
\end{align}
where we have used the well-known fact that one can use minus the conjugate momentum of a coordinate as a new Hamiltonian to reparametrize the motion via that given coordinate (see e.g. \citet{guckenheimer}). However, we can also use a ``pseudo-Newtonian" Hamiltonian of the form $H_\mathrm{pN}=(u_t^2-1)/2$ to obtain
\begin{equation}
H_\mathrm{pN}=\frac{1}{2}\sum_i (u_i)^2\,. \label{eq:flatpN}
\end{equation}
This Hamiltonian formally resembles the Hamiltonian of a free Newtonian particle and will give the correct motion of particles parametrized, however, by a pseudo-time $\tilde{t}$ given as
\begin{align}
\frac{\d u_i}{\d \tilde{t}} =- \frac{\partial H_\mathrm{pN}}{\partial x^i} = u_t \frac{\partial (-u_t)}{\partial x^i} = -u_t \frac{\d u_i}{\d t}\,,\label{eq:repar1}\\
\frac{\d x^i}{\d \tilde{t}} = \frac{\partial H_\mathrm{pN}}{\partial u^i} = -u_t \frac{\partial (-u_t)}{\partial u_i} = -u_t \frac{\d x^i}{\d t} \label{eq:repar2}\,.
\end{align}
So, we can conclude that this pseudo-time $\tilde{t}$ is related to the lab time on a particle-to-particle basis as $\d \tilde{t}=\d t/(-u_t)$ (for particles traveling forward in time, $-u_t$ is positive and has the meaning of specific energy of the particle). There is, however, a very important distinction between a truly Newtonian evolution of particles and this pseudo-Newtonian Ansatz; in Newtonian physics, the time parameter corresponding to $\tilde{t}$ is a globally valid lab-frame time coordinate; here, the parameter $\tilde{t}$ is valid only as a parameter along a single trajectory and cannot be directly tied to a global time coordinate.


\subsection{General stationary space-time} \label{subsec:general}


The point of this whole paper is to exploit the formal Ansatz discussed in the last Subsection in the following way. We find a Hamiltonian reproducing exactly the coordinate shapes of relativistic orbits which has a formally Newton-like form, albeit reparametrizing the orbits by some trajectory-specific pseudo-time. Then, we {\em postulate} this Ansatz Hamiltonian as a new pseudo-Newtonian Hamiltonian to be used in fully Newtonian calculations, where the pseudo-time $\tilde{t}$ is elevated to a globally valid time coordinate. This means that for {\em every} relativistic geodesic in the original space-time we will have some pseudo-Newtonian trajectory of identical coordinate shape, even though with a scrambled and rescaled time.

Consider a general stationary space-time with the metric $g_{ \mu\nu}$ and a set of coordinates in which the metric is stationary with respect to the coordinate $t=x^0$. This condition usually specifies the coordinate $t$ uniquely but otherwise our computations are covariant with respect to arbitrary coordinate transformations on the spatial hypersurface (coordinates $x^i$). We can then, analogously to the derivations above, find that the motion of particles parametrized by $t$ will be given by the Hamiltonian 
\begin{align}
\begin{split}
H_t &= -u_t \\ &= \omega^i u_i -\sqrt{(\omega^ i u_i)^2 - (g^{ij}u_i u_j +\kappa)/g^{00}} \label{eq:pt}\,,
\end{split}
\end{align}
where $\omega^i\equiv g^{0i}/g^{00}$, and we have also introduced the constant $\kappa$ to account for both massive $\kappa=1$ and massless particles $\kappa=0$. We can now again define a pseudo-Newtonian Hamiltonian $H_\mathrm{pN}=(u_t^2 - 1)/2$ to obtain
\begin{align}
 \begin{split}
H_\mathrm{pN} = &-\frac{1}{2}\frac{g^{ij}}{g^{00}} u_i u_j - \frac{1}{2}\left(\frac{\kappa}{g^{00}} + 1 \right) + \omega^i u_i (\omega^i u_i - \sqrt{\mathcal{D}})\,,
 \label{eq:HpN}
\end{split}
\end{align}
where $\mathcal{D} = (\omega^ i u_i)^2 - (g^{ij}u_i u_j +\kappa)/g^{00}$. By an identical derivation as in (\ref{eq:repar1}),(\ref{eq:repar2}), we obtain that this Hamiltonian generates trajectory evolution reparametrized by a pseudo-time $\tilde{t}$ such that $\d\tilde{t}=\d t/(-u_t)$. We now postulate this pseudo-Newtonian Hamiltonian as an effective Hamiltonian for Newtonian computations.

As can be verified by direct computation, the Hamiltonian (\ref{eq:HpN}) seamlessly reduces to the Hamiltonian (\ref{eq:flatpN}) in flat regions of the space-time. Furthermore, if we switch to SI units and use the weak-field metric $g^{ij} = \delta^{ij}(1 + 2 \Phi/c^2)\,,g^{00}=-(1-2\Phi/c^2)$, we obtain the Hamiltonian (\ref{eq:HpN}) to linear order in $c^{-2}$ as
\begin{align} 
H_\mathrm{pN} = \frac{1}{2}\left(1 + \frac{4 \Phi}{c^2}\right) \sum_i (u_i)^2 + \kappa\left(\Phi+\frac{2 \Phi^2}{c^2}\right)\,,
\end{align}
where we have shifted the Hamiltonian by a dynamically unimportant constant. In the case of massless particles ($\kappa=0$), we obtain the well known equations for the deviation of a light-ray in a gravitational field. For massive particles ($\kappa=1$), the zeroth order in $1/c^2$ gives simply the Newtonian Hamiltonian of a particle in a gravitational field, and the first $1/c^2$ order gives a post-Newtonian correction of first order. This gives us the confidence to call the Hamiltonian (\ref{eq:HpN}) a pseudo-Newtonian one.

We would like to use this opportunity to stress again that the idea of  {\em post}-Newtonian and {\em pseudo}-Newtonian descriptions is very different and the example above is probably the only point where a connection can be made. The post-Newtonian approximation is an iterative scheme reducing the error of computation at every order, whereas the pseudo-Newtonian Hamiltonian is in a sense always ``exact" with the reservation that it introduces a time-reparametrization as detailed in  (\ref{eq:repar1}),(\ref{eq:repar2}) (we discuss the implications in the next paragraph).


\subsection{Deviations from relativity}\label{subsec:dev}


The conclusion of this Section so far is that, provided that we give the same initial momenta $u_i$ as in the exact relativistic case, we are able to reproduce the exact shapes of relativistic trajectories via a fully Newtonian framework and the Hamiltonian (\ref{eq:HpN}). Two things will be different, however, and both stem from the fact that the Newtonian trajectory is reparametrized with respect to the relativistic one. 

First, the coordinate velocities at the same points of the trajectory will be different in the relativistic and the pseudo-Newtonian case due to the different time parametrization. Consider the following example: We want to compare whether we obtain the same orbit in the relativistic and pseudo-Newtonian description. Hence, we choose a coordinate point $x^i$ and a coordinate velocity $v^i$ as initial conditions for our comparison. Then, we evolve a particle with an initial condition $\d x^i/\d t=v^i$ in the relativistic case, and $\d x^i/\d\tilde{t}=v^i$ in the pseudo-Newtonian case. It is obvious that by this procedure we will obtain a particle on a {\em different} coordinate orbit in each case!

In this sense, the initial velocities leading to the same coordinate orbits are rescaled in the pseudo-Newtonian case by the total specific energy $-u_t$. On the other hand, the correspondence between the relativistic and pseudo-Newtonian case in terms of initial momenta and coordinate positions $u_i, \, x^i$ is always exact.

The second deviation of the pseudo-Newtonian case with respect to the relativistic one can be best illustrated on circular orbits. As follows from the previous discussion, if there is a set of circular orbits in the relativistic space-time with some (canonical) angular-momentum distribution, then this set of circular orbits along with the angular-momentum distribution will be exactly reproduced in the pseudo-Newtonian description.  However, the coordinate frequencies of these circular orbits will be deformed as 
\begin{align}
\Omega_\mathrm{pN} = \frac{\d \varphi}{\d \tilde{t}} = -u_t  \frac{\d \varphi}{\d t} = -u_t \Omega\,,\label{}
\end{align}
where $\Omega_\mathrm{pN}$ is the frequency along the orbit in the pseudo-Newtonian case, $ \Omega$ the original relativistic angular frequency, and $\varphi$ some angular coordinate. 

The relative error in the frequency $\eta_{\Omega}$can then be easily derived as
\begin{align}
\eta_\Omega\equiv \frac{\Omega_\mathrm{pN} - \Omega }{\Omega} = -(1 + u_t) \equiv -\mathcal{E}\,,\label{eq:frekerr}
\end{align}
where we have defined a new quantity $\mathcal{E}$ as the specific binding energy of the particle. (I.e., in an asymptotically flat space-time, $\mathcal{E}$ will be positive if the particle is bound and will represent the energy per unit mass needed to transport the particle to infinity.)

Another important deviation we obtain are different energies of the particles. Since the Hamiltonian (\ref{eq:HpN}) is conserved and reduces to Newtonian energy in weak fields, it is also natural to interpret it as a pseudo-Newtonian specific energy. In the convention we choose, $H_\mathrm{pN}$ is zero for a particle at rest in a flat part of the space-time and as such it is equal to minus the pseudo-Newtonian binding energy $H_\mathrm{pN}=-\mathcal{E}_\mathrm{pN}=(u_t^2 - 1)/2$. Then, we can easily derive that the relative error of the specific binding energy will be equal to
\begin{align}
\eta_\mathcal{E} \equiv \frac{\mathcal{E}_\mathrm{pN} - \mathcal{E}}{\mathcal{E}} =  -\frac{\mathcal{E}}{2} \label{eq:enerr}
\end{align}
For instance, in the case of the Schwarzschild space-time, the tightest bound circular orbit (with maximal $\mathcal{E}$) is the ISCO with $\mathcal{E}=1 - \sqrt{8/9} \approx 0.06$. I.e., in the Schwarzschild space-time the maximal error in binding energy and angular orbital frequency of circular orbits as predicted by the pseudo-Newtonian Hamiltonian (\ref{eq:HpN}) will be $3\%$ and $6\%$ respectively. 


\subsection{Massless particles}


We would also like to point out that in the special case of massless particles ($\kappa=0$), the Hamiltonian (\ref{eq:HpN}) can reproduce trajectories parametrized {\em exactly} by coordinate time $t$. 

The trick enabling us to do this lies in two facts. First, the shape of a null geodesic is completely insensitive to rescalings of four-velocity $u^\mu \to \lambda u^\mu$ where $\lambda$ is some constant. Second, if a vector $v^\mu$ satisfies four-velocity normalization for a massless particle $g^{\mu \nu}v_\mu v_\nu=0$, so does another vector $u^\mu = \lambda v^\mu$. 

As a result, we can always take an initial condition for the trajectory of a massless particle and rescale it so that $u_t=-1$ and thus $\d \tilde{t} = \d t$. Since $u_t$ is an integral of motion in stationary space-times, this property will be true along the whole trajectory and we will simply have $t=\tilde{t}$.


\section{Static space-times} \label{sec:timediag}


We now investigate the pseudo-Newtonian Hamiltonian from the previous Section in the class of metrics for which $g^{0i} =\omega^i= 0$. Considered along with the assumption of stationarity with respect to the time coordinate $t=x^0$, this class of space-times is easily recognized as {\em static space-times}. 

For these, equation (\ref{eq:HpN}) gives (we use the fact that in static metrics $g_{00}=1/g^{00}$)
\begin{equation}
H_{\mathrm{pN}}= - \frac{1}{2} g_{00}g^{ij} u_i u_j - \frac{\kappa}{2}(g_{00}+1). \label{eq:HpNd}
\end{equation}
In the case of static space-times, it is easy to execute a Legendre transform of the Hamiltonian (\ref{eq:HpNd}) (this is not possible to do in closed form for a general $\omega^i \neq 0$). We first obtain the relationship between momenta and pseudo-Newtonian velocities $\dot{x} \equiv \d x^i/\d \tilde{t} = \d x^{i}/\d t (1+\mathcal{E})$:
\begin{align}
&\dot{x}^i = \frac{\partial H}{\partial u_i} = - g_{00} g^{ij} u_i\,,\\
& u_i = -\frac{g_{ij}}{g_{00}} \dot{x}^j\,, \label{eq:pseudovel}
\end{align}
where we have used the fact that thanks to $g^{0i}=g_{0i} = 0$ the matrix $g_{ij}$ is the inverse of $g^{ij}$. It is then easy to see that the resulting Lagrangian $L = u_i \dot{x}^i - H$ reads
\begin{equation}
L_{\mathrm{pN}}= -\frac{1}{2}  \frac{g_{ij}}{g_{00}} \dot{x}^i \dot{x}^j  + \frac{\kappa}{2}(g_{00}+1)\,. \label{eq:LpN}
\end{equation}
It is also obvious from (\ref{eq:pseudovel}) that 
\begin{equation}
\frac{d x^i}{\d \tilde{t}} = (-u_t) \frac{\d x^i}{\d t} = -g_{00} \frac{d x^i}{\d \tau}\,,
\end{equation}
a fact we will use extensively in the analysis of fluid equations in Section \ref{sec:applications}.


\subsection{Geometrical interpretation of equations of motion}


The equations of motion corresponding to Lagrangian (\ref{eq:LpN}) can be put in a very elegant form
\begin{equation}
\ddot{x}^k = - \gamma^k_{\;jl} \dot{x}^j \dot{x}^l - \frac{\kappa}{2} \frac{g^{ik}}{g^{00}} g_{00,k}\,, \label{eq:diagmotion}
\end{equation}
where $\gamma^k_{\;jl}$ are the Christoffel symbols corresponding to the three-dimensional Riemannian metric $s_{ij}\equiv-g_{ij}/g_{00}$ known also as the optical or Fermat metric (see e.g. \citet{abralascart})
\begin{equation}
\gamma^k_{\;jl} = \frac{1}{2} s^{ki} \left( s_{ij,l} + s_{il,j} - s_{jl,i} \right)\,, \label{eq:pseudochr}
\end{equation}
where $s^{ij} = -g^{ij}/g^{00}$ is the inverse of $s_{ij}$. In other words, the motion of a relativistic massive particle ($\kappa=1$) in a static space-time can be, upon reparametrization, formulated as the motion of a geodesic in curved three-dimensional {\em space} in a potential field, and the motion of light ($\kappa=0$) corresponds simply to the motion of a geodesic in that deformed space. 

This notion has already been explored by \citet{AbraCurv} where the authors arrive to the same conclusion through fitting the Binet formula of a Newtonian particle in curved space so as to yield the same orbit shapes as in Schwarzschild space-time. Our work clarifies the general possibility of this ``shape reproduction" of orbits via the language of reparametrization.

We would like to point out the fact that even in the case of massive particles it is possible to describe their motion on the spatial hyperslice as a geodesic of a Riemannian metric. This metric is known as the Jacobi metric and it is energy dependent. The derivation of the Jacobi metric in static space-times and relation to previous results in the literature are discussed in Appendix \ref{app:jacobi}.


\subsection{Pseudo-Newtonian potentials}\label{subsec:pnpot}


Let us now interpret the Lagrangian (\ref{eq:LpN}) strictly as a Lagrangian of a Newtonian particle moving in Euclidean space. A part of the Lagrangian then must be the specific kinetic energy of the particle and the rest is a particular pseudo-gravitational potential. However, the Fermat metric $s_{ij}$ is generally {\em not} flat and we cannot interpret $s_{ij}\dot{x}^i \dot{x}^j/2$ as the specific kinetic energy of a particle in the flat Euclidean space of Newtonian physics!  As a consequence, a part of $s_{ij}$ must be absorbed into the potential, thus forming a ``generalized", velocity-dependent gravitational potential. 

The first step in identifying this velocity-dependent pseudo-Newtonian gravitational potential $\Phi_\mathrm{pN}$ is to interpret the coordinates in which we are working as some set of coordinates in Euclidean space. Then, using the Euclidean metric $d_{ij}$ in these coordinates we obtain the split of the Lagrangian as
\begin{align}
&L_{\mathrm{pN}}= \frac{1}{2}  d_{ij} \dot{x}^i \dot{x}^j - \Phi_{\mathrm{pN}}(x^i,\dot{x}^i),\; \label{eq:Lsplit}\\  &\Phi_{\mathrm{pN}} = -\frac{\kappa}{2}(g_{00}+1)  -\frac{1}{2} (s_{ij} - d_{ij}) \dot{x}^i \dot{x}^j .
\label{eq:VpN}
\end{align}
The part $s_{ij}-d_{ij}$ is then the ``non-flat deviation" of the Fermat metric inducing the extra effects which cannot be captured in a simple velocity-independent potential. 

In the case of an asymptotically flat space-time the pseudo-gravitational potential $\Phi_{\mathrm{pN}}$ goes asymptotically to zero if $g_{00} \to -1$ and $s_{ij}=-g_{ij}/g_{00} \to d_{ij}$. (An explicit example of $d_{ij}$ and $\Phi_\mathrm{pN}$ for the Schwarzschild space-time is given in the following Section \ref{sec:spher}.)


\section{Spherically symmetric space-times} \label{sec:spher}


The most prominent example to demonstrate the results of the last Section is the Schwarzschild metric. The formula for the pseudo-Newtonian Lagrangian (\ref{eq:LpN}) applied to the Schwarzschild space-time expressed in Schwarzschild coordinates $t,\,r,\,\vartheta,\,\varphi$ gives
 
\begin{equation}
L_{\mathrm{ TR}} = \frac{1}{2} \left( \frac{\dot{r}^2}{(1- 2M/r)^2} + \frac{r^2(\sin^2 \! \vartheta \,\dot{\varphi}^2 + \dot{\vartheta}^2)}{1-2M/r} \right) + \kappa \frac{M}{r}\,, \label{eq:LTR}
\end{equation}
which for $\kappa=1$ coincides with the Lagrangian derived from the equations of motion in the Schwarzschild space-time by \citet{tejeda1}. (The $\kappa=0$ case giving exact light-rays is proposed only here.) Similarly, one obtains the same formula as in \citet{tejeda2,sarkar} (eq. (\ref{eq:spherprev})) once applying formula (\ref{eq:LpN}), $\kappa=1$ to spherically symmetric space-times. 


\subsection{Extracting pseudo-Newtonian potentials} \label{subsec:potentials}


To obtain the pseudo-Newtonian potential $\Phi_\mathrm{pN}$ from the Lagrangian (\ref{eq:LTR}), we must first identify the ``natural metric'' $d_{ij}$. In the case of the Schwarzschild space-time in Schwarzschild coordinates this ``natural metric'' is of course the Euclidean metric in spherical coordinates $r,\,\vartheta,\,\varphi$, i.e.
\begin{equation}
d_{rr} = 1\, ,\; d_{\vartheta \vartheta} = r^2\, ,\;d_{\varphi \varphi} = r^2 \sin^2\! \vartheta\,. \label{eq:sphermet}
\end{equation}
This way the Tejeda-Rosswog Lagrangian reorganizes as follows (compare with equation (\ref{eq:Lsplit}), (\ref{eq:VpN}) and (\ref{eq:sphermet}))
\begin{equation}
 L_{\mathrm{ TR}} = \frac{1}{2} \left( \dot{r}^2 + r^2(\sin^2 \! \vartheta \,\dot{\varphi}^2 + \dot{\vartheta}^2) \right) - \Phi_\mathrm{pNS}\,,
\end{equation}
where
\begin{equation}
\begin{split} 
\Phi_\mathrm{pNS} = &-\kappa \frac{M}{r} -\frac{2 M (r-M)}{(r-2M)^2} \dot{r}^2 \\& + \frac{2M}{r-2M} r^2 (  \dot{\varphi}^2 \sin^2 \! \vartheta + \dot{\vartheta}^2)\,.
\end{split}
\end{equation}
This is also in concordance with the results in \citet{tejeda1}. 

However, we would also like to demonstrate that this ``split'' of the Lagrangian is not unique. Consider for instance the Schwarzschild metric expressed using the isotropic radius $R$ for which $r = R(1+M/2R)^2$
 
\begin{equation}
\d s^2 = -\left(\frac{1- \frac{M}{2R}}{1 + \frac{M}{2 R}}\right)^2 dt^2 + \left(1 + \frac{M}{2 R} \right)^4 \left(\d R^2 + R^2\d \Omega^2 \right)\,,
\end{equation}
where $\d \Omega^2 \equiv \d \vartheta^2  + \sin^2\! \vartheta \, \d \varphi^2$. From the perspective of these coordinates the ``natural flat metric'' is
\begin{equation}
d_{RR} = 1\, ,\; d_{\vartheta \vartheta} = R^2\, ,\;d_{\varphi \varphi} = R^2 \sin^2 \vartheta\,,
\end{equation}
which, in return, leads to the reorganization of the Tejeda-Rosswog Lagrangian as
\begin{equation}
 L_{\mathrm{ TR}} = \frac{1}{2} \left( \dot{R}^2 + R^2(\sin^2 \! \vartheta \,\dot{\varphi}^2 + \dot{\vartheta}^2) \right) - \Phi_\mathrm{pNI}\,,
\end{equation}
where
\begin{align}
\begin{split}
&\Phi_\mathrm{pNI} =  -\kappa \frac{4 M R}{\left(M + 2R \right)^2} \\&+ \left(\frac{16R^4(M-2R)^2}{(M+2R)^6}-1 \right) \left( \dot{R}^2 + R^2 (  \dot{\varphi}^2 \sin^2 \! \vartheta +\dot{\vartheta}^2) \right)\,.
\end{split}
\end{align}
Hence, the split into a ``usual Newtonian kinetic energy'' and the ``pseudo-Newtonian potential'' is conventional and relies heavily on what we think is the ``natural flat metric'' or the ``natural Newtonian interpretation of coordinates" in the curved space-time. Furthermore, the pseudo-Newtonian potentials cannot be simply combined with other external potentials because they are subject to the full non-linearity of relativistic source superposition. 

Nevertheless, as already mentioned, the pseudo-Newtonian Lagrangian (\ref{eq:LpN}) is, as a whole, in fact invariant with respect to transformations of the spatial coordinates ({\em not} with respect to transformations involving the time coordinate!). I.e., the whole Lagrangian $L_\mathrm{pN}$ is uniquely defined by the choice of the time coordinate and we will obtain covariantly the same physical behavior no matter which coordinate system or formal reorganization of the terms we use.


\section{Charged particles and perfect fluids} \label{sec:applications}


Since the development of a pseudo-Newtonian description is ultimately aimed at modeling a fluid in an accretion process, we now take the first steps towards a formulation of pseudo-Newtonian magneto-hydrodynamics. 

To do that, we first generalize the pseudo-Newtonian Hamiltonian to charged particles in electromagnetic fields in Subsection \ref{subsec:elmag}. Then, in Subsection \ref{subsec:fluid} we proceed to give hydrodynamic equations for a perfect fluid in the pseudo-Newtonian gravitational field. The inclusion of all the relevant physics to ultimately give a set of equations for e.g. pseudo-Newtonian radiative magneto-hydrodynamics near a black hole is out of the scope of the current paper.


\subsection{Charged particle motion}\label{subsec:elmag}


The relativistic Hamiltonian of the trajectory of a charged particle with specific charge $q$ in an electromagnetic field $A^\mu$ reads
\begin{equation}
H_{\mathrm{EM}\tau} = \frac{1}{2} g^{\mu \nu} (\pi_\mu - q A_\mu)(\pi_\nu - q A_\nu)\,,
\end{equation}
where $\pi_\mu = u_\mu + q A_\mu$ is canonically conjugate to $x^\mu$. Analogously to Section \ref{sec:pNHam} we invert the expression for the constant value of the Hamiltonian $H_{\mathrm{ EM}\tau} = -\kappa/2$ to get a Hamiltonian of an electrogeodesic parametrized by coordinate time
 
\begin{equation} 
\begin{split}
H_{\mathrm{EM}t} &= -\pi_t \\ & =  \omega^i u_i -\sqrt{(\omega^ i u_i)^2 - (g^{ij}u_i u_j +\kappa)/g^{00}}+q A_0\,, \label{eq:HEMt}
\end{split}
\end{equation}
with the important substitution $u_i = \pi_i - q A_i $.

Now it is easy to postulate the pN electromagnetic Hamiltonian $H_\mathrm{pNEM}$ as

\begin{equation}
H_\mathrm{pNEM} \equiv \frac{\pi_t^2 - 1}{2}\,, \label{eq:HpNEM}
\end{equation}
where we again have to assume the stationarity of the space-time metric $g_{ \mu \nu}$ with respect to $t=x^0$, but also stationarity of the electromagnetic potential $A_\mu$ because we would have problems with relating the $t$-dependence to the pseudo-time $\tilde{t}$-dependence of the field. We can formulate this assumption differently to make clear its gauge-dependence; we assume that the Maxwell tensor $F^{\mu \nu}$ is time-independent and we choose a gauge such that $A_\mu$ is also globally time-independent. 

The Hamiltonian (\ref{eq:HpNEM}) will, similarly to the Hamiltonian $H_\mathrm{pN}$ in equation (\ref{eq:HpN}), reproduce exact electrogeodesics parametrized by a new pseudo-time $\d \tilde{t} = -\d t/\pi_t $. Nevertheless, we do not give the explicit expression for $H_\mathrm{pNEM}$ in the general case because they are very long and can be easily evaluated using (\ref{eq:HEMt}) and (\ref{eq:HpNEM}). 

The only case in which $H_\mathrm{pNEM}$ reduces to an elegant expression with an easy Legendre transform is the case where the spacetime is static and the $A_0$ component of the electromagnetic field vanishes. I.e., for charged particle motion in static spacetimes with static magnetic fields we obtain the pseudo-Newtonian Hamiltonian
 
\begin{equation}
H_\mathrm{pNEM} = - \frac{1}{2} g_{00}g^{ij} (\pi_i - q A_i)(\pi_j - q A_j) - \frac{\kappa}{2}(g_{00}+1)\,,
\end{equation}
and the respective Lagrangian $L_{\mathrm{pNEM}}$ reads
\begin{equation}
L_{\mathrm{pNEM}}= -\frac{1}{2}  \frac{g_{ij}}{g_{00}} \dot{x}^i \dot{x}^j  + \frac{\kappa}{2}(g_{00}+1) + q A_j \dot{x}^j \,. \label{eq:LpNEM}
\end{equation}
That is, at least in this special case of static magnetic fields and static space-times the charged-particle dynamics can be obtained along the lines of the usual minimal coupling.


\subsection{Perfect-fluid equations} \label{subsec:fluid}


It is possible to derive pseudo-Newtonian fluid equations from first principles by starting from the Boltzmann equation governing the motion of particles on pseudo-Newtonian trajectories, and then finding its zeroth and first moment to obtain the continuity and Euler equation. We have tried this approach but it does not yield equations which fit well with their corresponding relativistic counterparts. 

Hence, we adopt an {\em ad hoc} approach where the relevant equations are derived as a ``pseudo-Newtonization'' of the exact relativistic equations. Furthermore, we restrict ourselves only to the case of static metrics, because as mentioned in Subsections \ref{subsec:general} and \ref{subsec:elmag} it is impossible to invert the pseudo-Newtonian equations of motion so as to feature explicitly the velocities rather than canonical momenta in the general case. 

 Consider the relativistic particle-conservation equation in coordinate time $t$
 \begin{equation}
\frac{\d n}{\d t}\bigg|_\mathrm{rel} = - \frac{n}{u^t \sqrt{-g}} \left[ \left( w^i u^t \sqrt{-g} \right)_{,i} + (u^t \sqrt{-g})_{,t} \right]\,,
 \end{equation} 
where $w^i = \d x^i/\d t$ is the coordinate velocity, $n$ the particle-number density, and $\d n/\d t = \partial n / \partial t + \partial n/ \partial x^i w^i$ the material derivative with respect to $t$. We will now need the following identity
\begin{equation}
w^i u^t \sqrt{-g} = v^i \sqrt{d}\,,
\end{equation}
where $v^i = \d x^i / \d \tilde{t}$ and $d = -\det(g_{ij})/g_{00}$. We can then reparametrize the continity equation using pseudo-time $\tilde{t}$ to obtain the {\em} exact particle-conservation equation as
\begin{equation}
\frac{\d n}{\d \tilde{t}}\bigg|_\mathrm{rel} = -\frac{n}{\sqrt{d}}\left[ \left(v^i \sqrt{d}\right)_{,i} + \left(u_t \sqrt{d} \right)_{,t} \right]\,.
\end{equation}
The term $\sim (u_t \sqrt{d} )_{,t}$ is a special-relativistic term which survives in flat space-time and spoils our otherwise very Newtonian form of the particle-conservation equation. Let us write it out explicitly
\begin{equation}
-\frac{n}{\sqrt{d}}\left(u_t \sqrt{d} \right)_{,t} = -\frac{n}{u_t} \frac{g_{ij}}{g_{00}} \frac{\partial v^i}{\partial t} v^j\,,
\end{equation}
which, in SI units, attains a factor $ c^{-2} $ relative to the other terms. If we then assume the velocities of the fluid and their variability are non-relativistic in the pseudo-Newtonian frame $v/c, v_{,t}/c \ll 1$, we can neglect this term. The assumption that $\d x^i/ \d \tilde{t}$ is small is to leading order equivalent to the assumption that $\d x^i/dt$ is small so this criterion can be also given in terms of the usual coordinate velocities.

Thus, we postulate the approximate, pseudo-Newtonian particle-conservation equation as
\begin{equation}
\frac{\d n}{\d \tilde{t}}\bigg|_\mathrm{pN} = -\frac{n}{\sqrt{d}} \left(v^i \sqrt{d}\right)_{,i}\,. \label{eq:PNcont}
\end{equation}
This pseudo-Newtonian equation will have a conserved particle number of the form
\begin{equation}
\mathcal{N} = \int n(x^i) \sqrt{d} \, \d^3 x\,,\; \frac{\d \mathcal{N}}{\d \tilde{t}}=0\,.
\end{equation}
Let us now consider the exact relativistic Euler equation in coordinate time in static space-times (see e.g. \citet{tejeda17} for the case of a general metric)
\begin{equation}
\begin{split}
\frac{\d^2 x^i}{\d t^2}\bigg|_\mathrm{rel} =
& -\left( \Gamma^i_{\;00} + \Gamma^i_{\;jk} w^j w^k \right)
\\
&- \frac{1}{(u^t)^2(\varepsilon + P)}\left( P_{,j} g^{ij} - P_{,t} g^{00} w^i \right)\,, \label{eq:RelEul}
\end{split}
\end{equation}
where $\varepsilon$ is the total energy density in the gas, and $P$ the pressure. We can again reparametrize this equation by $\tilde{t}$ to obtain
\begin{equation}
\begin{split}
\frac{\d^2 x^i}{\d \tilde{t}^2}\bigg|_\mathrm{rel} =& - \Gamma^i_{\;00}(u_t)^2 - \Gamma^i_{\;jk} v^j v^k + \frac{g_{00,k}}{g_{00}} v^k v^i 
 \\& - \frac{1}{(\varepsilon + P)}\left[(g_{00})^2 P_{,j} g^{ij} + P_{,t} u_t v^i \right]\,. \label{eq:ReparamRelEul}
\end{split}
\end{equation}
With the use of $(u_t)^2 = -g_{ij}v^i v^j/g_{00}- g_{00}$ we can reexpress the gravitational terms as
\begin{equation}
\begin{split}
&- \Gamma^i_{\;00}(u_t)^2 - \Gamma^i_{\;jk} v^j v^k + \frac{g_{00,k}}{g_{00}} v^k v^i =\\
&=- \frac{1}{2}s^{ij}g_{00,j} - \gamma^i_{\;jk} v^j v^k\,.
\end{split}
\end{equation}
I.e., as expected, the gravitational part of the acceleration of the fluid element is exactly equal to the acceleration of a pseudo-Newtonian particle (\ref{eq:diagmotion}). 

Since the gravitational part of (\ref{eq:ReparamRelEul}) is already pseudo-Newtonized, let us examine the hydrodynamic part. We assume 1) that the gas does not reach relativistic temperatures, in SI units $k_\mathrm{B} T \ll mc^2$, and 2) that the gas does not reach relativistic velocities in the pseudo-Newtonian frame, in SI units $v^2 \ll c^2$. Furthermore, we can estimate $P_{,t} \sim P_{,k} c_\mathrm{s}$ where $c_\mathrm{s}$ is the local sound speed (which is smaller than the speed of light). Hence, if we neglect terms from equation (\ref{eq:ReparamRelEul}) which are small in this approximation, we obtain the pseudo-Newtonian Euler equation as
\begin{equation}
\begin{split}
\frac{\d^2 x^i}{\d \tilde{t}^2}\bigg|_\mathrm{PN} =& - \frac{1}{2}s^{ij}g_{00,j} - \gamma^i_{\;jk} v^j v^k
 \\&- \frac{1}{\rho}(g_{00})^2 P_{,j} g^{ij}\,. \label{eq:PNEul}
\end{split}
\end{equation}
We can see that this Euler equation converges to the Newtonian limit automatically in the weak field without any additional special-relativistic terms. Additionally, it will reproduce most of the strong-field effects of its exact relativistic counter-part (\ref{eq:ReparamRelEul}).

To conclude, by using the reparametrization method and making assumptions which are reasonable for astrophysical applications, one is able to obtain a full set of pseudo-Newtonian fluid equations which are pseudo-Newtonian in the sense of automatically reducing to Newtonian equations in flat regions of space-time.

However, one must keep in mind the more subtle approximation introduced by the time-reparametrization. By evolving the elements of the fluid from some $\tilde{t}_0$ to some $\tilde{t}_0 + \delta \tilde{t}$ we are in fact evolving each element by $\delta t = -u_t \delta \tilde{t}$. This means that the elements in the pseudo-Newtonian evolution fall slightly out of sync as compared with the exact relativistic situation. This relative error does not show up in a single step but accumulates at a rate which is proportional to the relative differences of $-u_t$ between neighboring elements and also to the strength with which they interact. 

Hence, for the validity of the pseudo-Newtonian fluid evolution we also have to require that the length scale of the $-u_t$ variability is at all times much larger than the hydrodynamic interaction length-scale
\begin{equation}
\frac{(u_t)_{,i}}{u_t} \ll \frac{P_{,i}}{P}\,.
\end{equation}
In SI units and the weak-field limit, this criterion is to leading order in $c^{-1}$ 
\begin{equation}
\frac{1}{c^2} \left(\frac{1}{2}(v^2) + \Phi\right)_{,i} \ll \frac{P_{,i}}{P}\,,
\end{equation}
which is a criterion fulfilled in most physical applications.

For illustration and further applications, we have computed the pseudo-Newtonian Euler and particle-conservation equations explicitly in the Schwarzschild space-time in the usual radial Schwarzschild coordinates and Cartesian isotropic coordinates and include them in Appendix \ref{app:fleq}. The Cartesian-{\em isotropic} coordinates, based on the isotropic rather than the usual Schwarzschild radius, correspond to a pseudo-Newtonian metric which is isotropic at every space-time point. This set of coordinates could be useful in numerical schemes such as smoothed-particle hydrodynamics, because there the smoothing kernel of the particles can be entirely isotropic as long as the smoothing length is much shorter than the curvature scale of the space-time (compare e.g. with \citet{laguna1993smoothed}).

Of course, simply weighing the order of magnitude of the terms in the relativistic Euler and particle-conservation equations is not sufficient to fully asses the applicability of the set of pseudo-Newtonian fluid equations. However, the full investigation and testing of this set of equations by either numerical or analytical means is out of the scope of the current paper. Hence, one should understand equations (\ref{eq:PNcont}) and (\ref{eq:PNEul}) as a proposal whose usefulness can be shown by future work. On the other hand, we are confident that this set of equations will result in better results than naively implementing the gravitational accelerations (\ref{eq:diagmotion}) into Newtonian hydrodynamics without any other modification. 

Namely, in situations where densities and pressures are high enough to steer the motion of the fluid elements far away from free test-particle motion, we expect the additional strong-field coupling to the hydrodynamical degrees of freedom to become very important. For instance in the case of the circularized perfect-fluid equilibria in Schwarzschild space-time known as Polish doughnuts \citep{donut}, a simple computation shows that, up to some rescalings of density and angular momentum, our structural equations will yield the same structures of the doughnuts as the exact relativistic equations. If, however, we omit the $(g_{00})^2 g^{ij}$ factor in the pressure term in the Euler equation, one obtains radically different solutions such as equilibria extending to the horizon while being held by finite pressure gradients. Thus, including the strong-field factor in the pressure term is absolutely necessary for a good description of highly pressurized flows near the horizon.


\section{The Kerr space-time} \label{sec:kerr}


One of the most interesting goals in formulating pseudo-Newtonian frameworks is a satisfactory description of motion in the Kerr space-time at high values of the spin parameter $a$. By giving a useful description of highly spinning black holes one would extend the applicability of the pseudo-Newtonian framework to the vast majority of astrophysical black holes.  

The pseudo-Newtonian Hamiltonian given in Section \ref{sec:pNHam} will describe the motion in the Kerr space-time almost perfectly, and will lag with respect to the exact relativistic case only for large binding energy $\mathcal{E}=1+u_t$. However, giving a full description of motion of a fluid near a Kerr black hole has some difficulties as already described in Subsection \ref{subsec:fluid} and is thus out of the scope of the current paper. Hence, we will demonstrate the properties of the pseudo-Newtonian Hamiltonian in the Kerr space-time only on the motion of individual massive test particles and specifically on circular orbits and their close oscillations.

In Boyer-Lindquist coordinates $t,r,\vartheta,\varphi$ we have the non-zero inverse metric components of the Kerr metric (e.g. \citet{mtw,weinberg,grifpod})
\begin{equation}
\begin{split}
& g^{tt} =  -\frac{\mathcal{A}}{\Delta \Sigma} ,\\
& g^{rr} =   \frac{\Delta}{\Sigma}, g^{\vartheta \vartheta} = \frac{1}{\Sigma},\\
& g^{\varphi \varphi} =  \frac{\Delta-a^2\sin^2 \! \vartheta}{\Delta \Sigma \sin^2 \! \vartheta},\\
& g^{t \varphi} = \frac{2M r a}{\Delta \Sigma },
\end{split}
\end{equation} 
where $\Sigma = r^2 + a^2 \cos^2 \! \vartheta$, $\Delta = r^2 - 2Mr + a^2$ and $ \mathcal{A} = (r^2 + a^2)^2 -a^2 \Delta \sin^2 \! \vartheta $. The corresponding pseudo-Newtonian Hamiltonian (\ref{eq:HpN}) for massive particles $\kappa=1$ then reads
 
\begin{equation}
\begin{split}
H_\mathrm{pNK} = &\frac {1}{2 \mathcal{A}} \Big( \Delta^2 u_r^2 + \Delta u_\vartheta^2 + \frac{\Delta - a^2 \sin^2 \vartheta}{\sin^2 \vartheta}u_\varphi^2 \Big)  \\ &+ \frac {1}{2} \left( \frac{\Delta \Sigma}{\mathcal{A}} + 1  \right) + \omega u_\varphi (\omega u_\varphi - \sqrt{\mathcal{D}}), \label{eq:HpNK}
\end{split}
\end{equation}
 
where $\omega \equiv g^{t \varphi}/g^{tt} = -2Mra/\mathcal{A}$ and
 
\begin{equation}
\begin{split}
\mathcal{D} =& (\omega u_\varphi)^2 
 + \frac {1}{\mathcal{A}} \Big( \Delta^2 u_r^2 + \Delta u_\vartheta^2 + \frac{\Delta - a^2 \sin^2 \vartheta}{\sin^2 \vartheta}u_\varphi^2 \Big) + \frac{\Delta \Sigma}{\mathcal{A}}\,.
\end{split}
\end{equation} 
 
The Hamiltonian (\ref{eq:HpNK}) is formally very complicated. On the other hand, as described in Section \ref{sec:pNHam} and particularly Subsection \ref{subsec:dev}, $H_\mathrm{pNK}$ provides:
\begin{enumerate}
\item An angular momentum distribution over circular orbits exactly equal to the Kerr case,
\item absolutely exact behavior of the marginally bound ($H_\mathrm{pNK}=0,\,\mathcal{E}=0$) circular orbit as compared to the Kerr case including both the radius and rotation frequency, and 
\item an easily tractable upper error estimate for all bound circular orbits based on the binding energy of the ISCO and (\ref{eq:enerr}),(\ref{eq:frekerr}).  
\end{enumerate}

We will now discuss the errors induced to circular orbits and close oscillations around them. Furthermore, we will compare our Hamiltonian with the Lagrangian of \citet{ghosh} and offer a few remarks.


\subsection{Circular orbits} \label{subsec:circKerr}


The condition for a circular orbit is $\partial H_\mathrm{pNK}/\partial r =0$ with $u_r=u_\vartheta=0$, $\vartheta = \pi/2$ and some $r, u_\varphi$ to be determined. As described in Section \ref{sec:pNHam}, this condition will be fulfilled for exactly same pairs of $r,\, u_{\varphi}$ for $H_\mathrm{pNK}$ as for the exact relativistic case of a Kerr black hole. Hence, the formal angular momentum distribution $u_{\varphi\mathrm{c}}(r)$ over circular orbits is the same in both cases and reads \citep{bardeen}
\begin{equation}
u_{\varphi\mathrm{Kc}}(r) = \frac{\pm M^{1/2}(r^2 \mp 2aM^{1/2}r^{1/2} + a^2)}{\sqrt{r^3 - 3M r^{2} \pm 2aM^{1/2} r^{3/2}}}\,, \label{eq:angdistrib}
\end{equation} 
where the upper sign will always refer to the co-rotating circular orbits and the lower sign to the counter-rotating circular orbits. 

Using the formula for the pseudo-Newtonian Hamiltonian (\ref{eq:HpN}) and the definition of binding energy discussed in Subsection \ref{subsec:dev}, we can find the expression for pseudo-Newtonian binding energy in terms of the original relativistic one as  $\mathcal{E}_\mathrm{pN} = \mathcal{E}- \mathcal{E}^2/2$. That means that bound circular orbits will always have a lower energy in the pseudo-Newtonian case and, since the efficiency of accretion disks is estimated by the binding energy of the ISCO, the accretion disks in the pseudo-Newtonian fields will generally have lower efficiency than the ones in the corresponding relativistic space-times.

Also, since we know that all time rates such as the rotation frequency along a circular orbit will be rescaled by the factor $-u_t$ in the pseudo-Newtonian case and that $-u_t\in (0,1)$ for bound orbits, then the pseudo-Newtonian rotation frequencies of {\em bound} circular orbits will always be smaller than in the exact Kerr case. On the other hand, from the marginally bound orbit inwards to the black hole, we have {\em unbound} orbits, $-u_t > 1$, and there the pseudo-Newtonian frequencies will be {\em higher} than in the exact Kerr case. 

Furthermore, as we approach the photon sphere the energy of circular orbits diverges and both the time and energy scales become vastly different from the relativistic case. Hence, for accretion disc modeling, we must cut off the dynamics somewhere between the radius of the marginally bound orbit and the photon orbit.

Let us now give a few more explicit expressions for the behavior of the circular orbits. The specific binding energy $\mathcal{E}_\mathrm{Kc}$ of circular orbits in the relativistic case reads
\begin{equation}
\mathcal{E}_\mathrm{Kc}(r) = 1-\frac{r^{3/2} - 2Mr^{1/2}\pm aM^{1/2}}{\sqrt{r^3 - 3M r^{2} \pm 2aM^{1/2} r^{3/2}}}\,.
\end{equation} 
On the other hand, the specific binding energy $\mathcal{E}_\mathrm{pNKc}$ of circular orbits in the pseudo-Newtonian case yields
 
\begin{equation}
\mathcal{E}_\mathrm{pNKc}(r) = \frac{1}{2} \left[ 1- \frac{(r^{3/2} - 2Mr^{1/2}\pm aM^{1/2})^2}{r^3 - 3M r^{2} \pm 2aM^{1/2} r^{3/2}}  \right]\,.
\end{equation} 
 
The position of the marginally bound circular orbit is given by solving $\mathcal{E}_\mathrm{Kc}= \mathcal{E}_\mathrm{pNKc}=0$, which gives
\begin{equation}
r_\mathrm{Kmb} = 2M - a +2\sqrt{M(M-a)}\,.
\end{equation}
I.e., for the circular orbit of radius $r_\mathrm{Kmb}$ the correspondence between the pseudo-Newtonian and relativistic case is perfect both in energy and frequency.
  
The angular rotation frequency $\Omega_\mathrm{Kc}\equiv\d \varphi/\d t$ of circular orbits in the relativistic case reads
\begin{equation}
\Omega_\mathrm{Kc}(r) = \frac{\pm M^{1/2}}{r^{3/2} \pm a M^{1/2}}\,.
\end{equation}
In the pseudo-Newtonian case we just have the analogous angular rotation frequency given as $\Omega_\mathrm{pNKc} \equiv \d\varphi/\d \tilde{t} = (1-\mathcal{E}_\mathrm{Kc}) \Omega_\mathrm{Kc}$ which gives
 
\begin{equation}
\Omega_\mathrm{pNKc} (r) = \frac{\pm M^{1/2} (r^{3/2} - 2Mr^{1/2}\pm aM^{1/2})}{(r^{3/2} \pm a M^{1/2})\sqrt{r^3 - 3M r^{2} \pm 2aM^{1/2} r^{3/2}}}\,.
\end{equation}
 
Using the expressions above, we can easily plot the properties of the circular orbits in both models for a black hole with any spin $a$ and compare them. 


\subsection{Innermost stable circular orbit} \label{subsec:ISCOKerr}


The most important estimate of accretion disc behavior comes from studying the properties of the ISCO. For instance, the binding energy of the ISCO is equal to the efficiency of the accretion process in a radiatively efficient thin accretion disc \citep{abramowiczlivrev}. In other cases, the orbital frequency of the ISCO is proposed to distinguish between black-hole candidates and neutron stars \citep{psaltis}. Also, as has been already mentioned, the binding energy of the ISCO gives an upper error estimate for various deviations of the pseudo-Newtonian description from the relativistic case and can thus serve as an overall indicator of the applicability of the pseudo-Newtonian Hamiltonian (\ref{eq:HpNK}).

The radius of the innermost stable circular orbit is \citep{bardeen}
 
\begin{align}
& r_\mathrm{ISCO} = M \left[ 3 + Z_2 - \sqrt{(2-Z_1)(4+Z_1 + 2 Z_2)} \right]\,,\\
& Z_1 = (1-a^2/M^2) \left[(1-a^2/M^2) + (1+a^2/M^2) \right]\,,\\
& Z_2 = \sqrt{3a^2/M^2 +1+ Z_1^2}\,.
\end{align}
 
\begin{figure}
\begin{center}
\includegraphics[width=0.48\textwidth]{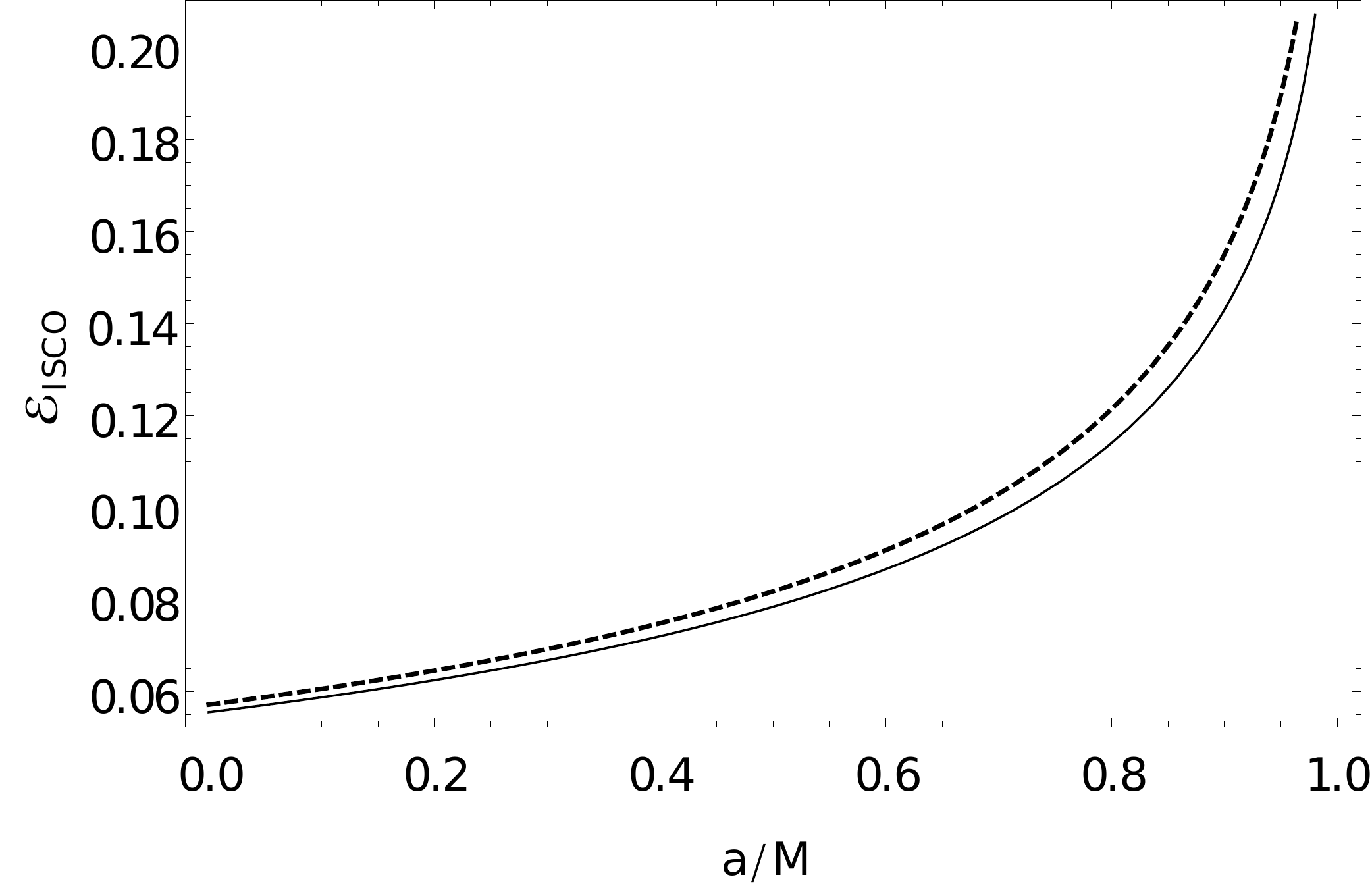}
\caption{Specific binding energy of the ISCO $\mathcal{E}_\mathrm{ISCO}$ in the Kerr space-time (dashed) and the pseudo-Newtonian counter-part (full line). } \label{fig:Eisco}
\end{center}
\end{figure}

\begin{figure}
\begin{center}
\includegraphics[width=0.48\textwidth]{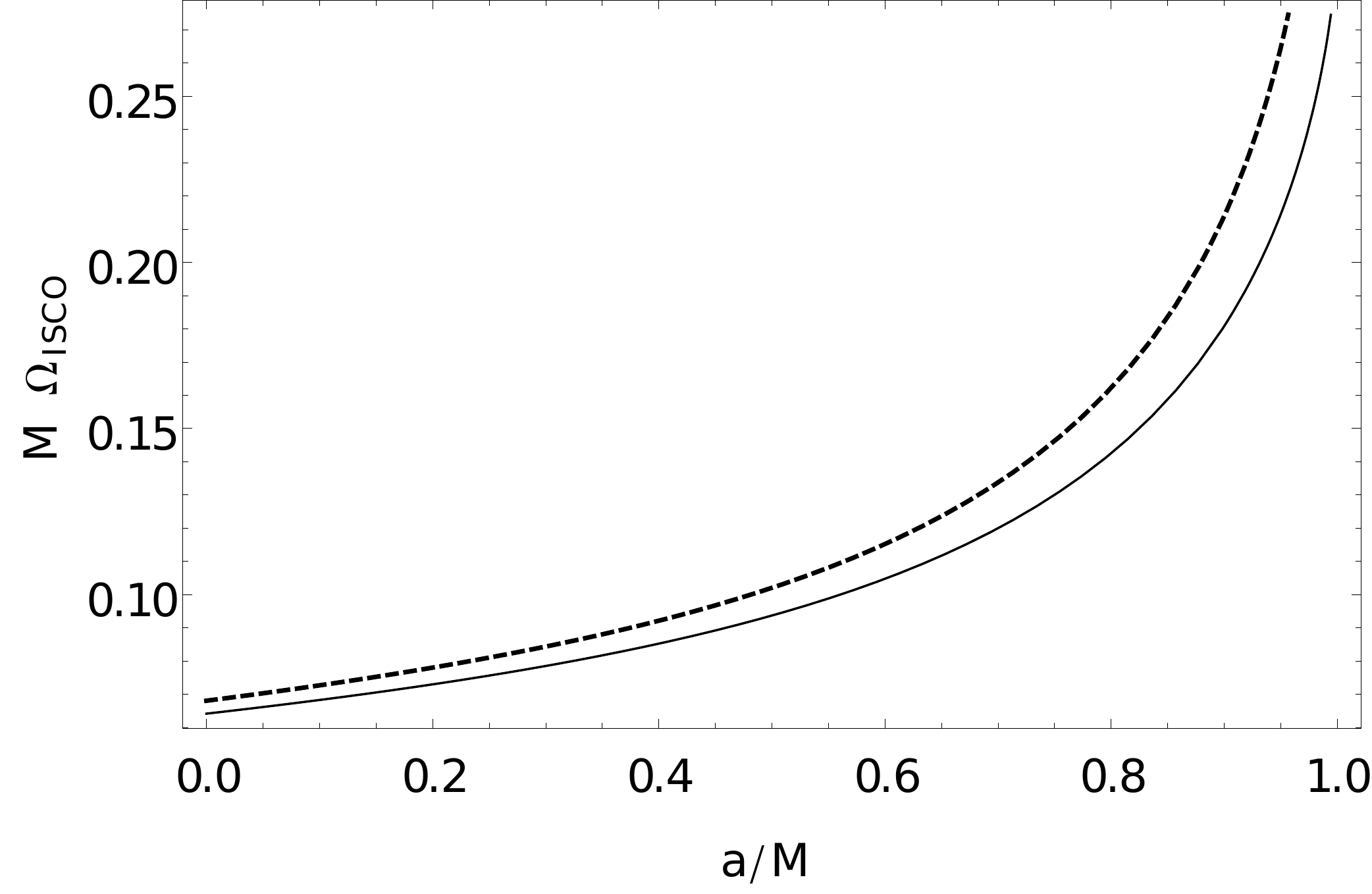}
\caption{Angular rotation frequency $\Omega_\mathrm{ISCO}$ of the ISCO in the Kerr space-time (dashed) and the pseudo-Newtonian counter-part (full line). The frequency is given in units of $M^{-1}$ ($ G^{-1} M^{-1} c^3$ in SI units) so that the result is scalable with respect to the mass of the black hole.} \label{fig:OMisco}
\end{center}
\end{figure}

It is quite obvious that the substitution of $r_\mathrm{ISCO}$ into the expressions for energy or frequency gives very complicated formulas. Hence, we only compare the relations $\mathcal{E}_\mathrm{ISCO}(a)$ in Figure \ref{fig:Eisco} and the angular frequencies of the ISCO in Fig. \ref{fig:OMisco}.

As we go to higher spins of the black hole, the relativistic binding energy $\mathcal{E}_\mathrm{ISCO}$ grows and thus also the relative errors in the pseudo-Newtonian values of binding energy and frequency of circular orbits. For instance, if we want the relative error of the binding energy of the ISCO to be less than $10\%$ in the pseudo-Newtonian model, we can only use the Hamiltonian (\ref{eq:HpNK}) for spins $a<0.96M$, which is a reasonable bound. However, if we set the tolerance in the relative error of ISCO binding energy to $5\%$, we can only use the pseudo-Newtonian Hamiltonian for spins $a<0.68M$, which is rather restrictive. Either way, it is not very reasonable to use the pseudo-Newtonian Hamiltonian all the way up to the extremal $a=M$ black holes because there we have the relativistic ISCO binding energy $\mathcal{E}_\mathrm{ISCO} = 1-1/\sqrt{3}\approx 0.42$, and thus the relative error of the binding energy of the ISCO about $21\%$.

\subsection{Small perturbations of circular orbits}


Let $\delta r, \delta u_r, \delta \vartheta, \delta u_\vartheta$ be small deviations from a circular orbit. At the point of reflectional symmetry $\vartheta=\pi/2$ all the first $\partial/\partial \vartheta$ derivatives of the Hamiltonian vanish. Additionally for $u_\vartheta=0$, the first $\partial/\partial u_\vartheta$ derivatives are also zero. Hence, the linearized equations of motion for the small deviations decouple into two sectors corresponding to the purely radial (epicyclic) and purely vertical oscillations around the circular orbit. The equations for the purely radial oscillations in matrix form read
\begin{equation}
\begin{pmatrix}
 0 & -\frac{\partial^2 H_\mathrm{pNK}}{ \partial r^2} \\[0.4em]
  \frac{\partial^2 H_\mathrm{pNK}}{\partial u_r^2 } & 0 \\
 \end{pmatrix} \begin{pmatrix}
  \delta u_r \\[0.4em]
  \delta r \\
 \end{pmatrix} = 
 \begin{pmatrix}
  \delta \dot{u}_r \\[0.4em]
  \delta \dot{r} \\
 \end{pmatrix}.
\end{equation}
Because we are considering perturbations around $u_r=0$, the diagonal terms corresponding to first $\partial/\partial u_r$ derivatives of the Hamiltonian are also zero. The solution for $\delta u_r,\,\delta r$ is an oscillating solution with a frequency
\begin{equation}
\omega_{r\mathrm{pNK}} = \left( \frac{\partial^2 H_\mathrm{pNK}}{\partial u_r^2 } \frac{\partial^2 H_\mathrm{pNK}}{ \partial r^2} \right)^{1/2}, \label{eq:omr}
\end{equation}
where the expression is evaluated at $\vartheta=\mathrm{\pi}/2,\, u_r=u_\vartheta =0,\, u_\varphi = u_{\varphi\mathrm{c}}(r)$. Similarly for the purely vertical oscillations we obtain the vertical oscillation frequency
\begin{equation}
\omega_{ \vartheta \mathrm{pNK}} = \left( \frac{\partial^2 H_\mathrm{pNK}}{\partial u_\vartheta^2 } \frac{\partial^2 H_\mathrm{pNK}}{ \partial \vartheta^2} \right)^{1/2}. \label{eq:omth}
\end{equation}
Expressions (\ref{eq:omr}) and (\ref{eq:omth}) along with the substitution of $u_{\varphi \mathrm{c}}$ from eq. (\ref{eq:angdistrib}) give the pseudo-Newtonian oscillation frequencies as cumbersome analytical expressions.

\begin{figure*}
\begin{center}
\includegraphics[width=0.48\textwidth]{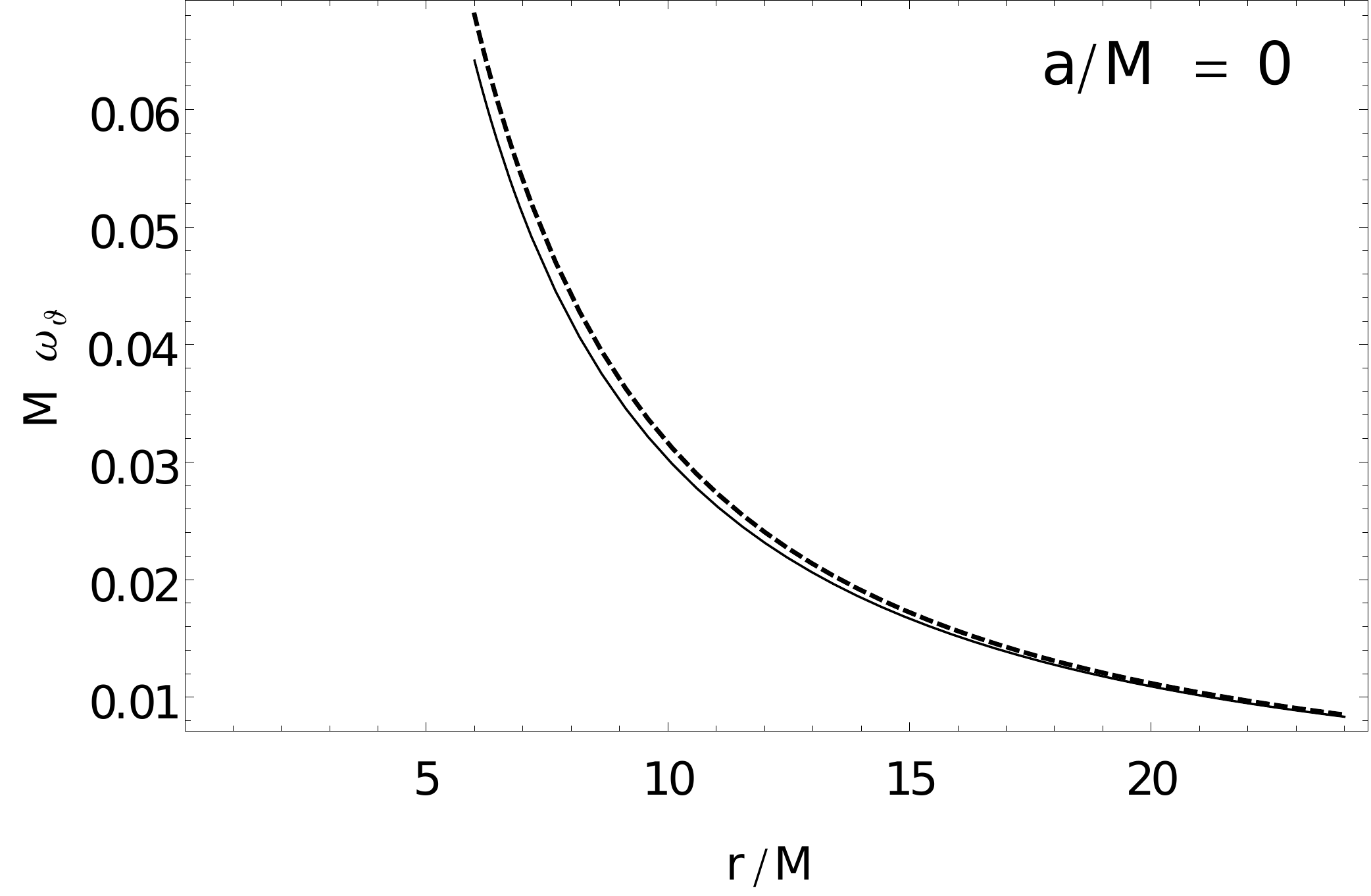} \includegraphics[width=0.48\textwidth]{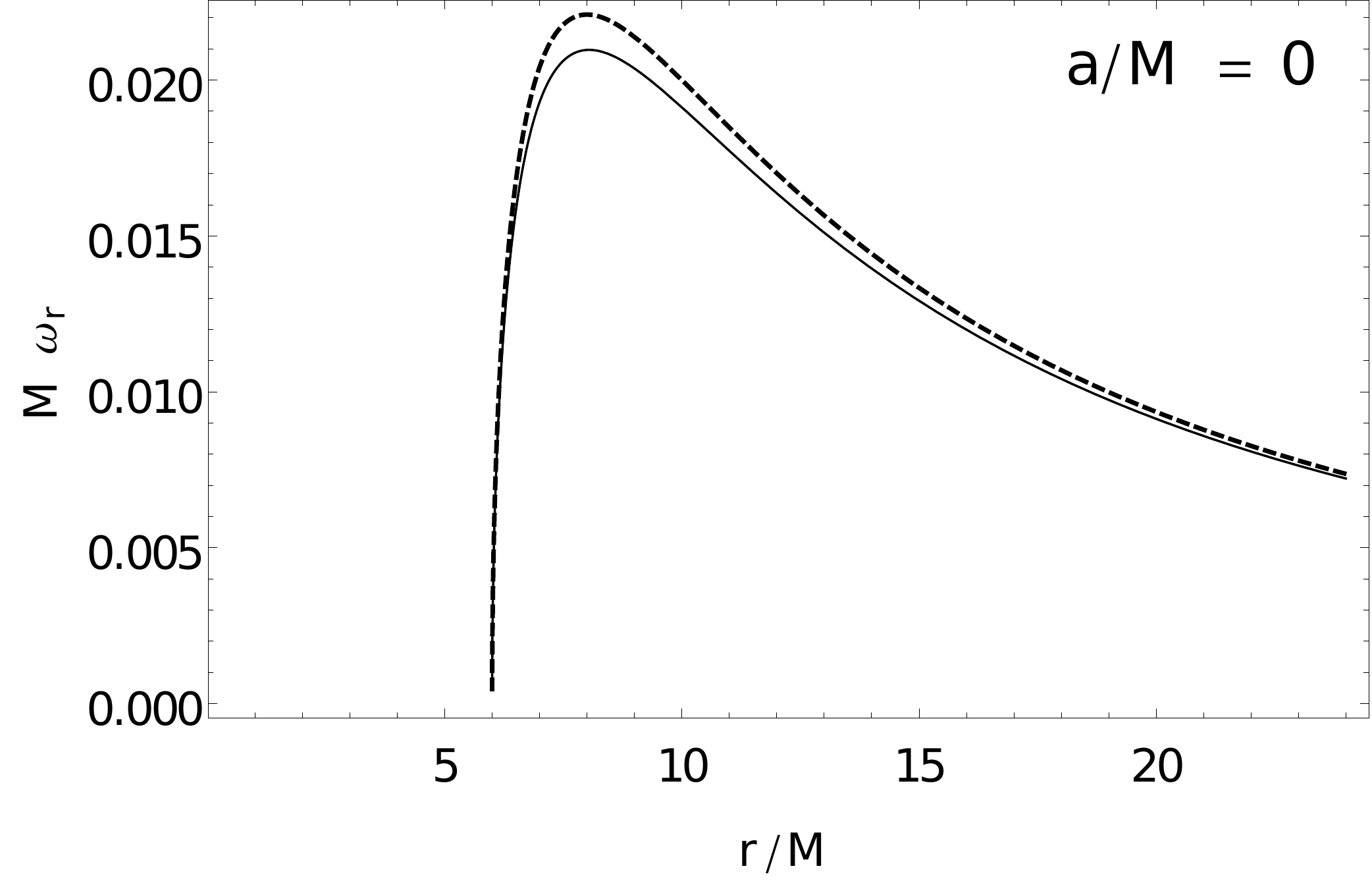} \includegraphics[width=0.48\textwidth]{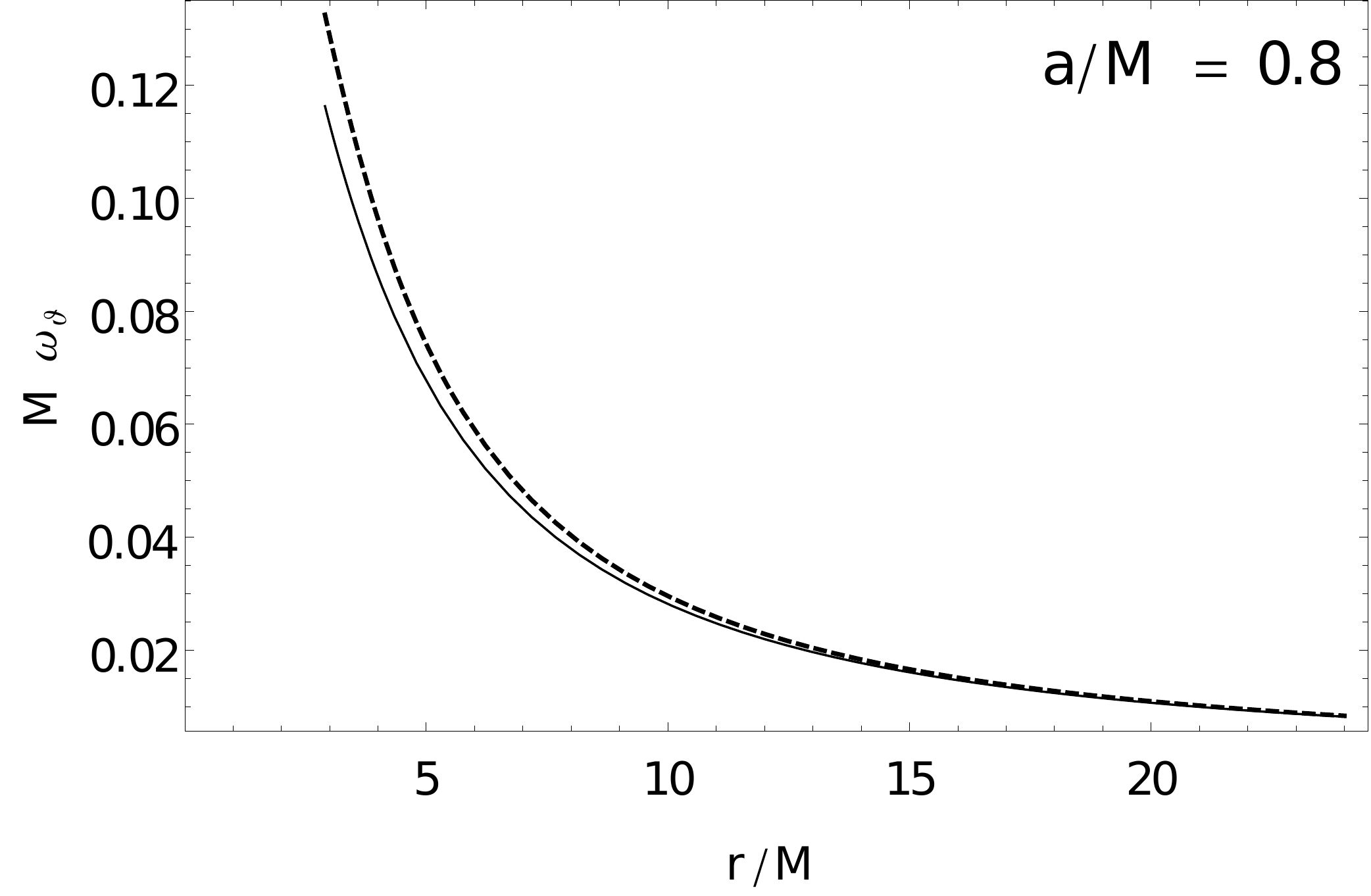} \includegraphics[width=0.48\textwidth]{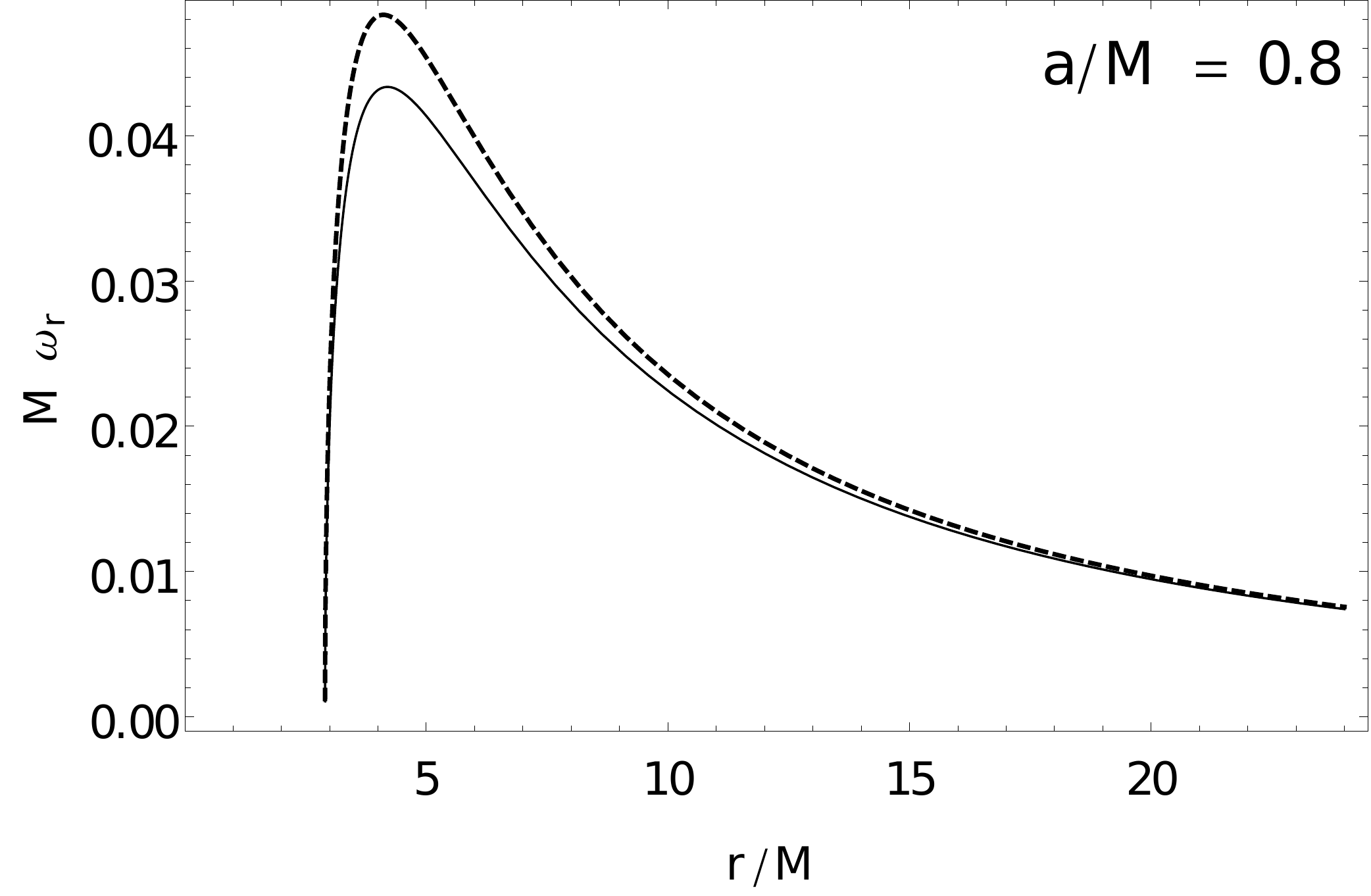} 
 \includegraphics[width=0.48\textwidth]{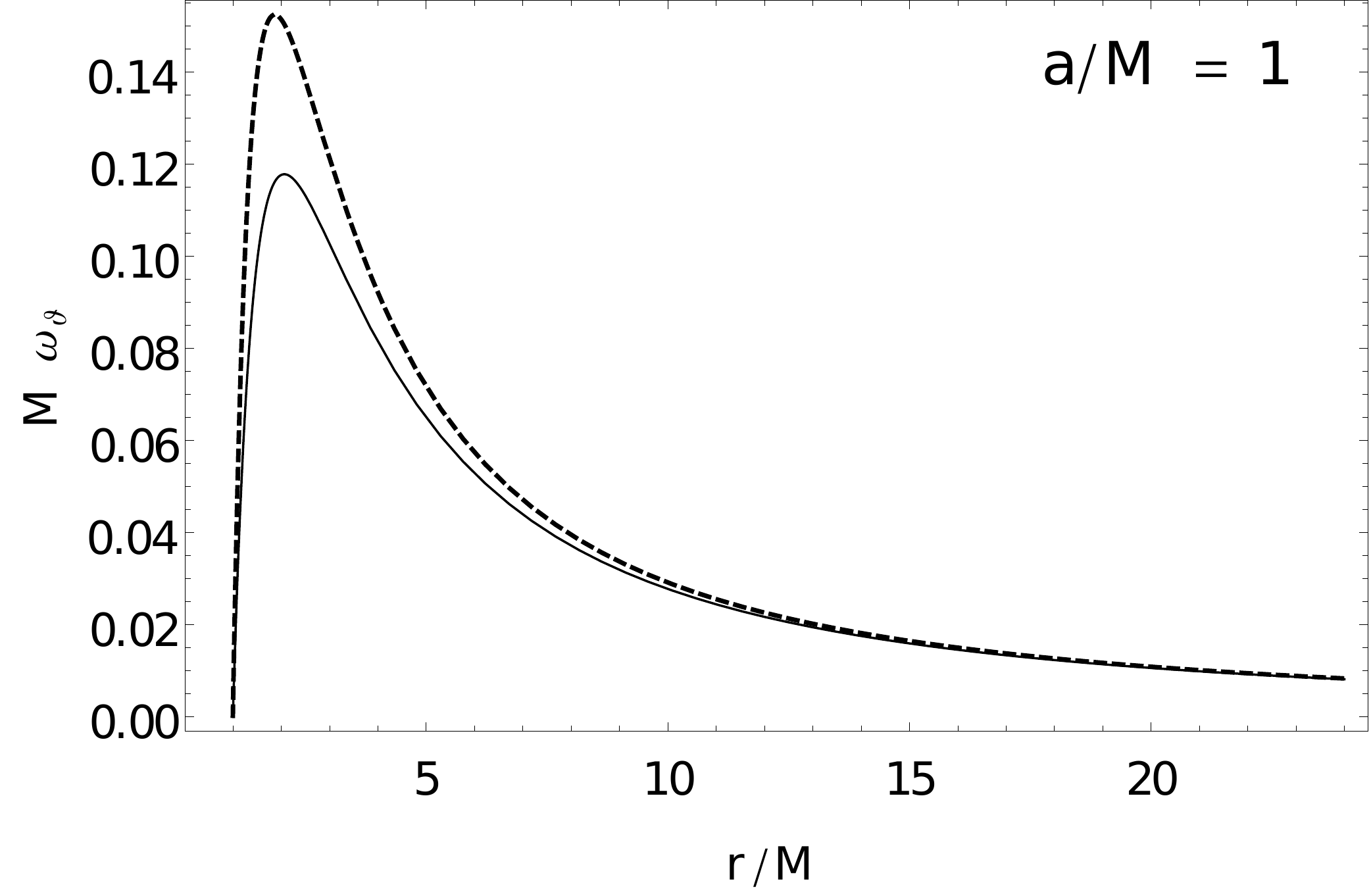} \includegraphics[width=0.48\textwidth]{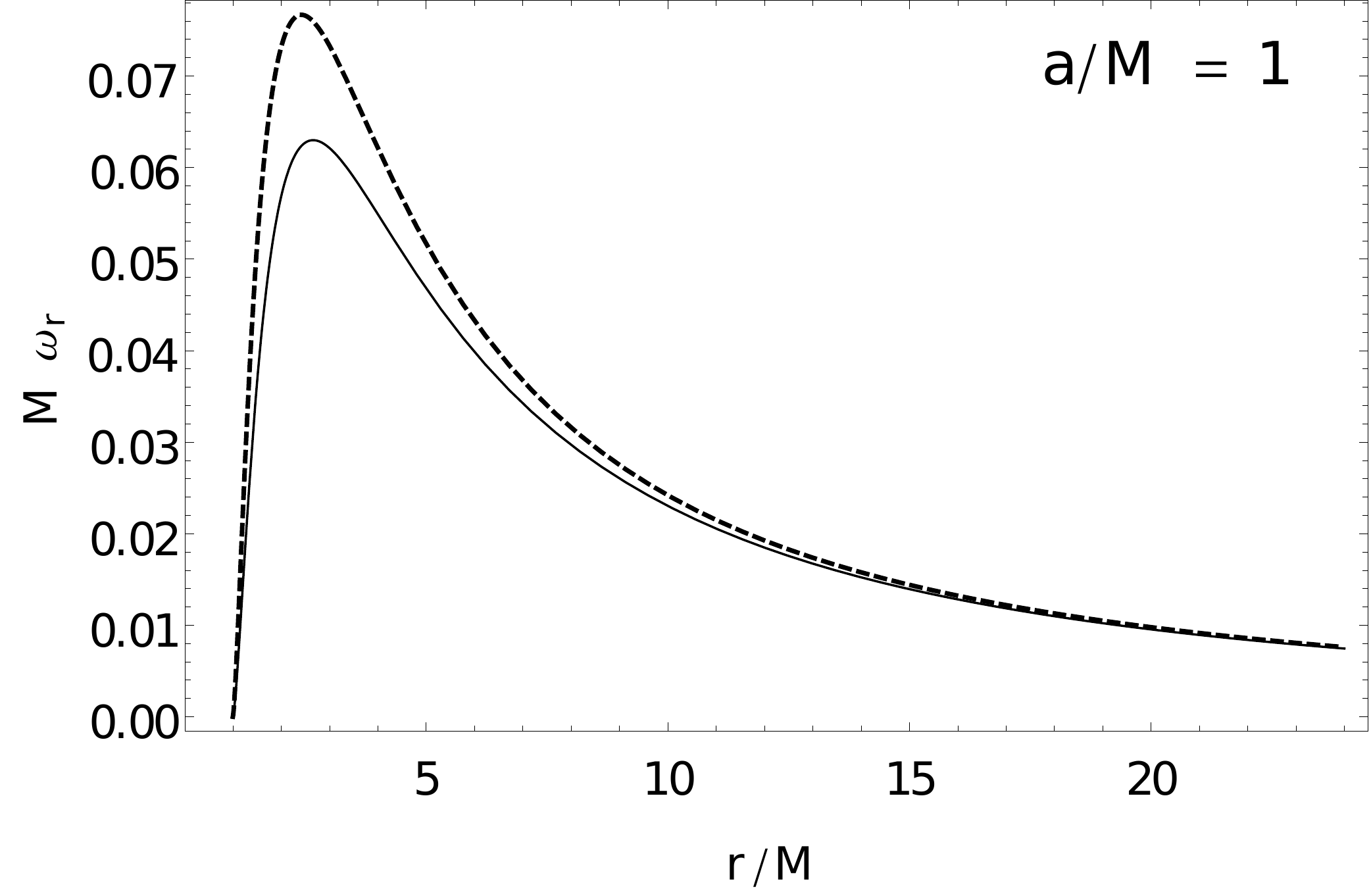} 
\caption{Comparison of vertical (left) and radial (right) oscillation frequencies of perturbed circular orbits between the exact Kerr values (dashed) and the values obtained with the pseudo-Newtonian Hamiltonian (\ref{eq:HpNK}) (full). The $r$ plot ranges are fixed for all plots while the $\omega$-ranges are individually adjusted and only values up to the ISCO are plotted. It is not entirely obvious that the highest relative errors between the frequencies is always in the left-most part of the individual plots (at the ISCO) but we have verified this fact both by analytical and numerical means. } \label{fig:oscil}
\end{center}
\end{figure*}

Nevertheless, we can find a workaround by considering that the pseudo-Newtonian Hamiltonian $H_\mathrm{pNK}$ is equal to $H_\mathrm{pNK} = (u_t^2 - 1)/2 = (H_t^2 - 1)/2$, where $H_t$ is the Hamiltonian generating exact relativistic motion parametrized by coordinate time. The frequencies are thus given as
\begin{align}
&\omega_{r\mathrm{pNK}} = -u_t \left( \frac{\partial^2 H_t}{\partial u_r^2 } \frac{\partial^2 H_t}{ \partial r^2} \right)^{1/2} = -u_t \omega_{r}\,,\\
&\omega_{ \vartheta \mathrm{pNK}} = -u_t\left( \frac{\partial^2 H_t}{\partial u_\vartheta^2 } \frac{\partial^2 H_t}{ \partial \vartheta^2} \right)^{1/2} = -u_t \omega_{ \vartheta }\,,
\end{align}
where $\omega_{r},\, \omega_{ \vartheta}$ are the exact oscillation frequencies in Kerr space-time. However, these  are well known (see e.g. \citet{abramowiczlivrev})
\begin{align}
&\omega_{r} = \Omega_\mathrm{Kc} \sqrt{1-6 r M^{-1} + 8 a r^{-3/2}M^{1/2} -3a^2 r^{-2}} \,, \\
&\omega_{ \vartheta} = \Omega_\mathrm{Kc} \sqrt{1-4 a r^{-3/2}M^{1/2} +3a^2 r^{-2}} \,.
\end{align}
As a result, we see that once again the relative difference between the relativistic and pseudo-Newtonian frequencies will be equal to the binding energy of the circular orbit
\begin{equation}
\eta_{\omega \alpha} \equiv \frac{\omega_{\alpha\mathrm{pNK}} - \omega_{\alpha}}{\omega_{\alpha\mathrm{pNK}}} = \mathcal{E}_\mathrm{c}\,,
\end{equation}
where $\alpha=r,\, \vartheta$. This means, in particular, that the maximum error in the oscillation frequencies will be once again given by the binding energy of the ISCO as given in Fig. \ref{fig:Eisco}. Hence, the pseudo-Newtonian model is useful in a similar range as discussed in the previous Subsection \ref{subsec:ISCOKerr} even for accretion models where the disc oscillations are relevant. For illustration, we plot the oscillation frequencies for a number of values of the spin parameter in Fig. \ref{fig:oscil}.

\subsection{Remarks on the Ghosh-Sarkar-Bhadra Lagrangian} \label{subsec:ghoshremarks}


\citet{ghosh} derived a Lagrangian (the GSB Lagrangian) for the motion of test particles in the equatorial plane which naturally offers itself for comparison with the herein presented Hamiltonian (\ref{eq:HpNK}). We point out the differences of the approach of \citet{ghosh} and a few non-trivial facts about the GSB Lagrangian.

Instead of covariant velocity components $u_i$, the GSB Lagrangian is constructed by a series of Ansatzes using the contravariant (canonically {\em non}-conjugate) components $u^i$. As a consequence, the dynamics are restricted only to the equatorial plane (the pseudo-Kerr Hamiltonian (\ref{eq:HpNK}) presented in this paper applies to any $\vartheta$) and it seems that there is no simple characterization of the GSB Lagrangian in terms of reparametrized geodesics.

The Ghosh-Sarkar-Bhadra Lagrangian reads
 
\begin{equation}
\begin{split}
L_{\mathrm{ GSB}} =  &\frac{1}{2(r-2M)^2 (1 + \gamma \dot{\varphi})} \left(\frac{r^3 (r-2M)}{\Delta} \dot{r}^2 + \Delta r^2 \dot{\varphi}^2 \right) \\&+ \frac{M}{r}(1- \gamma \dot{\varphi}),
\end{split} \label{eq:LGSB}
\end{equation}
where $\gamma = 2Ma/(r-2M)$. It is interesting that the GSB Lagrangian (\ref{eq:LGSB}) has an impractical {\em Hamiltonian} counterpart, whereas the herein presented pseudo-Kerr Hamiltonian (\ref{eq:HpNK}) has a {\em Lagrangian} counterpart complicated beyond usefulness. Complications associated with either the forward or backward Legendre transform seem to be a general feature of Lagrangians and Hamiltonians modeling the gravitomagnetic effects in the Kerr space-time.

The GSB Lagrangian has certain problems with the angular momentum distribution of circular orbits connected to the singularities of the effective potential. For the angular momenta of circular orbits $\lambda_{\mathrm{ GSBc}}$ it holds that
\begin{equation}
\begin{split}
&\lambda_{\mathrm{ GSB c}} = \frac{-Q \pm \sqrt{Q^2-4 R}}{2},\\ \label{eq:lGSB}
&Q = \frac{4 a^3r M - 6 Ma r (r^2 + a^2)}{a^2 r(r-2M) - r(r-3M)(r^2 + a^2)},\\
&R = \frac{M(r^2 + a^2)[r(r^3 + 3a^2)-2a^2 r]}{a^2 r(r-2M) - r(r-3M)(r^2 + a^2)}.
\end{split}
\end{equation}
What was not clearly stated or shown in the original paper is the fact that this angular momentum distribution has a singularity at a radius $r_\mathrm{s}$ given by
\begin{equation}
a^2 (r_{\mathrm{ s}}-2M) - (r_{\mathrm{ s}}-3M)(r_{\mathrm{ s}}^2 + a^2) = 0, \label{eq:GSBsing}
\end{equation}
for which the solution varies quite uniformly from $r_{\mathrm{ s}}=3M$ for $a=0$ to $r_{\mathrm{ s}} \approx 3.1M$ for $a=M$. 

Even though the authors state that the marginally bound ($\mathcal{E}=0$) circular orbit exists up to $a \approx 0.7M$ and that the potential is thus useful up to such values, there is a possible issue with the marginally bound orbit; the angular momentum distribution (\ref{eq:lGSB}) crosses the singularity (\ref{eq:GSBsing}) before reaching the radius of the marginally bound orbit already for $a \gtrsim 0.45 M$. Amongst other things, this means that the Keplerian circular orbits have a ``singular pause'' before reaching the marginally bound orbit and the matter density of a stationary accretion disc would almost certainly exhibit non-physical behavior at the singular $r=r_{\mathrm{ s}}$.

Hence, the GSB Lagrangian should be considered as useful only for $r\gtrsim 3.1M$, and if the marginally bound orbit is important in the given model, only $a \lesssim 0.45 M$ should be considered. The point where even the ISCO collides with this singularity is $a \approx 0.7M$ (which is the reason why the authors of \citep{ghosh} were not able to find the ISCO beyond that spin). Amongst other things, this means that for a near-Keplerian accretion disk near a black hole with spin $a \approx 0.7M$ the singularity (\ref{eq:GSBsing}) is very near its edge and exotic effects might ensue. Thus, it seems commendable to use the Lagrangian (\ref{eq:LGSB}) only for spins well below $a \approx 0.7 M$.

 
\section{Conclusion}

 
We have developed and studied a generalized pseudo-Newtonian formalism appropriate for particles, light, and fluids in stationary space-times. In the case of static, spherically symmetric space-times, our formalism coincides with previous results in the literature \citep{tejeda1,tejeda2,sarkar}. In general static space-times (without gravitomagnetic terms in the metric), this formalism has elegant geometric interpretations and allows for the full explicit development of fluid equations which can be understood as a particular kinematic limit of the fully relativistic equations. Additionally, we have included electromagnetic fields influencing the motion in the case of particles with charge. 

{\bf\em As already stated in Subsection \ref{subsec:fluid}, the presented pseudo-Newtonian fluid equations should be further investigated by comparing their numerical and analytical solutions with relativistic counterparts so that their proper applicability is fully understood. The undeniable point of our analysis is that a naive implementation of the pseudo-Newtonian acceleration of individual particles into a Newtonian code necessarily neglects further coupling of the strong gravitational field to the hydrodynamical degrees of freedom and may lead to pathological behavior of the fluid near the black hole horizon.}

The herein presented pseudo-Newtonian framework is exceptional in its clear and direct derivation from the original space-time which we want to mimic. As a result, we have obtained explicit bounds on the various errors our pseudo-Newtonian formalism introduces. The general conclusion is that the various relative errors in this description grow linearly with the specific binding energy of the motion in question. 

This has consequences for the applicability of Newtonian numerical codes used along with the pseudo-Newtonian equations of motion. We have derived that for stationary structures near non-rotating black holes we can expect the result of a pseudo-Newtonian computation to be accurate within a few percent of relative error in temporal and energetic quantities, and this error then nonlinearly grows with the spin of the black hole up to a few tens of percent.  

One particular shortcoming of the formalism is the fact that in space-times which are stationary but not static (with dragging or gravitomagnetic terms in the metric), it is not possible to express particle Lagrangians in a closed form and we are constrained to Hamiltonian formalism. Similarly, the corresponding fluid equations would be expressed in some set of canonical momenta rather than velocities, which is incompatible with common numerical codes. This means that pseudo-Newtonian fluid evolution near a spinning black hole probably has to be approached in a different manner.

It is easy to include electromagnetic terms into the pseudo-Newtonian Euler equation (\ref{eq:PNEul}), albeit only in the non-physical case when 1) the fluid is composed exclusively of charged particles of a single value of charge and the current thus proportional to the velocity of the fluid, and 2) the currents in the fluid have no backreaction the static, externally imposed fields. In the actuallly physically relevant case of a quasi-neutral fluid with the current deviating from the velocity and back-reacting to the electromagnetic field, the equations become very complicated, and the separation of strong-field, ``kinematically non-relativistic'', and Newtonian terms becomes much more subtle. Hence, we leave the question of pseudo-Newtonian magneto-hydrodynamics as a possibility for future work.


\acknowledgments
We would like to thank Emilio Tejeda, Old{\v r}ich Semer{\'a}k, and Volker Perlick for useful discussions on the preliminary versions of the paper. We would also like to thank the anonymous reviewer for many useful remarks on the paper.
VW is grateful for support from grants GAUK-2000314 and SVV-260211 and a Ph.D. grant of the German Research Foundation within its Research Training Group 1620 {\em Models of Gravity}.


\appendix


\section{Jacobi metric in static space-times}\label{app:jacobi}


It is a well known result in the theory of classical mechanics that for a time-independent Lagrangian given in the form
\begin{equation}
L = \frac{1}{2} d_{ij} \dot{x}^i \dot{x}^j - V(x)\,, \label{eq:Lclass}
\end{equation}
one can define an energy-dependent metric called the Jacobi metric as
\begin{equation}
j_{ij} = (E - V)d_{ij}\,, 
\end{equation}
where $E=d_{ij} \dot{x}^i \dot{x}^j/2 + V$ is the energy integral of motion.  Then if $E>V$, the trajectories corresponding to the Lagrangian (\ref{eq:Lclass}) on a fixed hypersurface $E=\mathrm{const.}$ will be, up to a reparametrization, geodesics corresponding to the metric $j_{ij}$ (see e.g. \citet{pettini} for more details).
 Hence, following this pattern we identify $s_{ij}$ as the analogy of $d_{ij}$, $\kappa(g_{00}+1)/2$ as $-V$, and the value of the Hamiltonian $\tilde{E}=H_\mathrm{pN}=((u_t)^2 -1)/2$ as the energy integral. When the dust settles, we obtain the Jacobi metric 
\begin{equation}
j_{ij} = -\frac{(u_t)^2 + \kappa g_{00}}{2} \frac{g_{ij}}{g_{00}} \,. \label{eq:jacobimetric}
\end{equation}
I.e., on constant $u_t$ ($\tilde{E}$) hypersurfaces and in static space-times, the full four-dimensional geodesics can always be described as three-dimensional geodesics on the spatial hypersurface with the Jacobi metric (\ref{eq:jacobimetric}).  This result has been recently given by \citet{gibbons} by considering the action of a geodesic in static space-times; the formalism here provides a connection between the result of \citet{gibbons} and the usual notion of the Jacobi metric known from classical mechanics.


\section{Evolution equations for a perfect fluid near a black hole in various coordinates} \label{app:fleq}



\subsection{Cartesian isotropic coordinates}


The most simple Cartesian-like expression of the Schwarzschild metric is given by a transformation to the isotropic radial coordinate $r = (M+2R)^2/(4 R)$  (introduced already in Subsection \ref{subsec:potentials}, for a thorough discussion of the coordinates see \citet{mtw}) and then transforming to a set of coordinates $x^{(1)},\, x^{(2)},\, x^{(3)}$
\begin{align}
& x^{(1)}= R \sin \vartheta \cos \varphi\, , \\
& x^{(2)} = R \sin \vartheta \sin \varphi \, , \\
& x^{(3)}= R \cos \vartheta \, , \\
& R = \sqrt{\sum_{i=1}^3 \left(x^{(i)}\right)^2}\,.
\end{align}
The metric then takes the form
 
\begin{equation}
\d s^2 =  -\left(\frac{2R- M}{2R+ M}\right)^2 \d t^2 + \frac{(2R + M)^4}{16R^4} \sum_{i=1}^3 \left(\d x^{(i)}\right)^2\,.
\end{equation}
I.e., the isotropic coordinates regularize the spatial part of the metric at the horizon while preserving the $g_{0i} = g^{0i}=0$ structure of the metric essential to the elegance of the pseudo-Newtonian formalism. Other sets of ``horizon-penetrating" coordinates which make even the temporal part of the metric regular at the horizon exist (see for example \citet{fonthorizon}), but they necessarily violate the $g_{0i} = g^{0i}=0$ condition. That is, the pseudo-Newtonian metric $s_{ij} \equiv g_{ij}/g_{00}$ will always be singular at the horizon and the singularity can only be ``softened" by a set of coordinates such as the isotropic ones.

Either way, we know from the discussion in Subsection \ref{subsec:dev} that circularized motion at the location of the photon orbit will always have very large errors as compared to the exact relativistic case, so we should beware extending the simulation up to there for circularized flows. On the other hand, circular motion at the marginally bound orbit is expected to exactly reproduce the relativistic features. Thus, we recommend to cut off a simulation of an accretion flow somewhere between the marginally bound circular orbit $r_\mathrm{mb}=4M \to R_\mathrm{mb}= (3+2\sqrt{2})M/2$ and the photon sphere $r_\mathrm{ps}=3M \to R_\mathrm{ps}=(2+\sqrt{3})M/2$. 

The pseudo-Newtonian metric $s_{(i)(j)} = -g_{(i)(j)}/g_{00}$ obtains the form
\begin{equation}
s_{(i)(j)} = \frac{(2R + M)^6}{16R^4 (2R- M)^2} \delta_{ij} \,, \label{eq:Sdef}
\end{equation}
where $\delta_{ij}$ is the Cronecker delta. The pseudo-Newtonian Christoffel symbols corresponding to this metric then have this simple form
\begin{align}
& \gamma^{(i)}_{\;\; (j)(k)} = -\frac{2M}{R^2 (4R + M)} (\delta_{ij} x^{(k)} + \delta_{ik} x^{(j)} - \delta_{jk} x^{(i)})\,.
\end{align}
We can thus easily express the respective pseudo-Newtonian equations of motion for a single particle (\ref{eq:diagmotion}) as
\begin{equation}
\ddot{x}^{(i)} = \frac{4M}{R^2 (4R + M)} \sum_{k=1}^3 ( \dot{x}^{(i)}\dot{x}^{(k)}x^{(k)} - \frac{1}{2}\dot{x}^{(k)}\dot{x}^{(k)}x^{(i)}) - \frac{64 M (2R -M)^3 R^3}{(2R +M)^9} x^{(i)} \,.
\end{equation}
The factor figuring in the particle-conservation equation then is 
\begin{equation}
\sqrt{d} = \frac{(2R+M)^7}{64 R^6 (2R-M)}
\end{equation}
and the particle-conservation equation reads

\begin{equation}
\dot{n} = -n \sum_{i=1}^3 \left( \frac{\partial v^{(i)}}{\partial x^{(i)}} + \frac{2 M (3 M - 8 R)}{R^2 (2 R - M) (2R +M)} x^{(i)} v^{(i)} \right)
\,,\label{eq:contIso}
\end{equation}
which corresponds to a conserved total particle number of the form
\begin{equation}
\mathcal{N} = \int n(x^{(i)})  \frac{(2R+M)^7}{64 R^6 (2R-M)} \d^3 x\,.
\end{equation}
The Euler equation for a fluid differs from the single-particle acceleration only by the $-P_{,j} g^{ij} g_{00}^2/\rho$ term and thus takes the form
\begin{equation}
\begin{split}
\ddot{x}^{(i)} = 
& \frac{4M}{R^2 (4R + M)} \sum_{k=1}^3 ( \dot{x}^{(i)}\dot{x}^{(k)}x^{(k)} - \frac{1}{2}\dot{x}^{(k)}\dot{x}^{(k)}x^{(i)}) - \frac{64 M (2R -M)^3 R^3}{(2R +M)^9} x^{(i)}
\\
& -\frac{P_{,(i)}}{\rho}  \frac{16 R^4(2R-M)^4}{(2R + M)^8}\,. \label{eq:eulerIso}
\end{split}
\end{equation}


\subsection{Schwarzschild radial coordinates}


In Schwarzschild coordinates $r, \varphi, \vartheta$ the non-zero components of the metric $s_{ij}$ read
\begin{equation}
s_{rr} = \frac{1}{(1-2M/r)^2}\,,\quad s_{\vartheta \vartheta} = \frac{r^2 }{1-2M/r}\,,\quad  s_{\varphi \varphi} = \frac{r^2 \sin^2 \! \vartheta}{1-2M/r}\,.
\end{equation}
From that we compute the non-zero coefficients $\gamma^i_{\;jk}$ as
\begin{align}
&\gamma^r_{\;rr} = -\frac{2 M}{r(r-2M)}\,, \quad \gamma^r_{\;\vartheta \vartheta} = -(r-3M)\,,\quad \gamma^r_{\;\varphi \varphi} = -(r-3M)\sin^2 \! \vartheta \,, 
\\
&\gamma^\varphi_{\; \varphi r} = \gamma^\varphi_{\; r \varphi} = \frac{r-3M}{r(r-2M)}\,, \quad \gamma^\varphi_{\; \varphi \vartheta} = \gamma^\varphi_{\; \vartheta \varphi} = \cot \vartheta\,,
\\
&\gamma^\vartheta_{\; \vartheta r} = \gamma^\vartheta_{\; r \vartheta } = \frac{r-3M}{r(r-2M)}\,, \quad \gamma^\vartheta_{\; \varphi \varphi} = - \sin \vartheta \cos \vartheta \,.
\end{align}
The corresponding gravitational accelerations of individual particles are then easy to find by direct computation or in \citet{tejeda1}. 

The volume density factor is $\sqrt{d} = r^2 \sin \vartheta/(1-2M/r)$ and the particle-conservation equation reads
\begin{equation}
\dot{n} = - n  \left( \frac{\partial \dot{r}}{\partial r}  +  \frac{\partial \dot{\vartheta}}{\partial \vartheta} +  \frac{\partial \dot{\varphi}}{\partial \varphi} + \frac{2(r-3M)}{r(r-2M)} \dot{r} + \cot \vartheta \, \dot{\vartheta}\right)\,,
\end{equation}
which corresponds to a conserved total particle number
\begin{equation}
\mathcal{N} = \int_0^\infty \!\!\! \int_0^\pi\!\! \int_0^{2\pi}\!\! \frac{n\, r^2 \! \sin \vartheta}{1 - 2M/r } \,\d r\, \d \vartheta\, \d \varphi \,.
\end{equation}
The Euler equations in Schwarzschild coordinates then read
\begin{align}
\ddot{r} &= \frac{2M}{r(r-2M)}\dot{r}^2 + (r-3M)[\dot{\vartheta}^2 + \sin^2\! \vartheta \, \dot{\varphi}^2]  -\frac{1}{\rho}\left(1 - \frac{2M}{r}\right)^3\frac{\partial P}{\partial r}  - \frac{M}{r^2} \left( 1 - \frac{2M}{r} \right)^2 \,,
\\[0.8em] 
\ddot{\varphi} &= - \frac{2(r-3M)}{r(r-2M)} \dot{r} \dot{\varphi} - 2\cot \vartheta \, \dot{\vartheta} \dot{\varphi}  + -\frac{1}{\rho\, r^2 \sin^2 \! \vartheta}\left(1 - \frac{2M}{r}\right)^2\frac{\partial P}{\partial \varphi}
\\[0.8em] 
\ddot{\vartheta}  &=  - \frac{2(r-3M)}{r(r-2M)} \dot{r} \dot{\vartheta} - 2 \sin \vartheta \cos \vartheta \,  \dot{\varphi}^2 -\frac{1}{\rho r^2}\left(1 - \frac{2M}{r}\right)^2\frac{\partial P}{\partial \vartheta} \,, 
\end{align}
Following the pattern of this Appendix, one should be able to derive the fluid equations in any set of coordinates.


\bibliography{literatura}

\begin{thebibliography}{34}%
\makeatletter
\providecommand \@ifxundefined [1]{%
 \@ifx{#1\undefined}
}%
\providecommand \@ifnum [1]{%
 \ifnum #1\expandafter \@firstoftwo
 \else \expandafter \@secondoftwo
 \fi
}%
\providecommand \@ifx [1]{%
 \ifx #1\expandafter \@firstoftwo
 \else \expandafter \@secondoftwo
 \fi
}%
\providecommand \natexlab [1]{#1}%
\providecommand \enquote  [1]{``#1''}%
\providecommand \bibnamefont  [1]{#1}%
\providecommand \bibfnamefont [1]{#1}%
\providecommand \citenamefont [1]{#1}%
\providecommand \href@noop [0]{\@secondoftwo}%
\providecommand \href [0]{\begingroup \@sanitize@url \@href}%
\providecommand \@href[1]{\@@startlink{#1}\@@href}%
\providecommand \@@href[1]{\endgroup#1\@@endlink}%
\providecommand \@sanitize@url [0]{\catcode `\\12\catcode `\$12\catcode
  `\&12\catcode `\#12\catcode `\^12\catcode `\_12\catcode `\%12\relax}%
\providecommand \@@startlink[1]{}%
\providecommand \@@endlink[0]{}%
\providecommand \url  [0]{\begingroup\@sanitize@url \@url }%
\providecommand \@url [1]{\endgroup\@href {#1}{\urlprefix }}%
\providecommand \urlprefix  [0]{URL }%
\providecommand \Eprint [0]{\href }%
\providecommand \doibase [0]{http://dx.doi.org/}%
\providecommand \selectlanguage [0]{\@gobble}%
\providecommand \bibinfo  [0]{\@secondoftwo}%
\providecommand \bibfield  [0]{\@secondoftwo}%
\providecommand \translation [1]{[#1]}%
\providecommand \BibitemOpen [0]{}%
\providecommand \bibitemStop [0]{}%
\providecommand \bibitemNoStop [0]{.\EOS\space}%
\providecommand \EOS [0]{\spacefactor3000\relax}%
\providecommand \BibitemShut  [1]{\csname bibitem#1\endcsname}%
\let\auto@bib@innerbib\@empty
\bibitem [{\citenamefont {Einstein}()}]{einstein}%
  \BibitemOpen
  \bibfield  {author} {\bibinfo {author} {\bibfnamefont {A.}~\bibnamefont
  {Einstein}},\ }\href@noop {} {\bibinfo  {journal} {Albert Einstein:
  Akademie-Vortr{\"a}ge: Sitzungsberichte der Preu{\ss}ischen Akademie der
  Wissenschaften 1914-1932}\ ,\ \bibinfo {pages} {78}}\BibitemShut {NoStop}%
\bibitem [{\citenamefont {Blanchet}(2006)}]{blanchetlivrev}%
  \BibitemOpen
\bibfield  {journal} {  }\bibfield  {author} {\bibinfo {author} {\bibfnamefont
  {L.}~\bibnamefont {Blanchet}},\ }\href@noop {} {\bibfield  {journal}
  {\bibinfo  {journal} {Living Rev. Relat.}\ }\textbf {\bibinfo {volume} {9}},\
  \bibinfo {pages} {4} (\bibinfo {year} {2006})}\BibitemShut {NoStop}%
\bibitem [{\citenamefont {Sasaki}\ and\ \citenamefont
  {Tagoshi}(2003)}]{sasakilivrev}%
  \BibitemOpen
  \bibfield  {author} {\bibinfo {author} {\bibfnamefont {M.}~\bibnamefont
  {Sasaki}}\ and\ \bibinfo {author} {\bibfnamefont {H.}~\bibnamefont
  {Tagoshi}},\ }\href@noop {} {\bibfield  {journal} {\bibinfo  {journal}
  {Living Rev. Relat.}\ }\textbf {\bibinfo {volume} {6}} (\bibinfo {year}
  {2003})}\BibitemShut {NoStop}%
\bibitem [{\citenamefont {Abramowicz}\ and\ \citenamefont
  {Fragile}(2013)}]{abramowiczlivrev}%
  \BibitemOpen
  \bibfield  {author} {\bibinfo {author} {\bibfnamefont {M.~A.}\ \bibnamefont
  {Abramowicz}}\ and\ \bibinfo {author} {\bibfnamefont {P.~C.}\ \bibnamefont
  {Fragile}},\ }\href@noop {} {\bibfield  {journal} {\bibinfo  {journal}
  {Living Rev. Relat.}\ }\textbf {\bibinfo {volume} {16}},\ \bibinfo {pages}
  {1} (\bibinfo {year} {2013})}\BibitemShut {NoStop}%
\bibitem [{\citenamefont {Paczy{\'n}sky}\ and\ \citenamefont
  {Wiita}(1980)}]{paczynsky}%
  \BibitemOpen
  \bibfield  {author} {\bibinfo {author} {\bibfnamefont {B.}~\bibnamefont
  {Paczy{\'n}sky}}\ and\ \bibinfo {author} {\bibfnamefont {P.~J.}\ \bibnamefont
  {Wiita}},\ }\href@noop {} {\bibfield  {journal} {\bibinfo  {journal} {A\&A}\
  }\textbf {\bibinfo {volume} {88}},\ \bibinfo {pages} {23} (\bibinfo {year}
  {1980})}\BibitemShut {NoStop}%
\bibitem [{\citenamefont {Abramowicz}(2009)}]{abramowiczPW}%
  \BibitemOpen
  \bibfield  {author} {\bibinfo {author} {\bibfnamefont {M.~A.}\ \bibnamefont
  {Abramowicz}},\ }\href@noop {} {\bibfield  {journal} {\bibinfo  {journal}
  {A\&A}\ }\textbf {\bibinfo {volume} {500}},\ \bibinfo {pages} {213} (\bibinfo
  {year} {2009})}\BibitemShut {NoStop}%
\bibitem [{\citenamefont {Tejeda}\ and\ \citenamefont
  {Rosswog}(2013)}]{tejeda1}%
  \BibitemOpen
  \bibfield  {author} {\bibinfo {author} {\bibfnamefont {E.}~\bibnamefont
  {Tejeda}}\ and\ \bibinfo {author} {\bibfnamefont {S.}~\bibnamefont
  {Rosswog}},\ }\href@noop {} {\bibfield  {journal} {\bibinfo  {journal}
  {MNRAS}\ }\textbf {\bibinfo {volume} {433}},\ \bibinfo {pages} {1930}
  (\bibinfo {year} {2013})}\BibitemShut {NoStop}%
\bibitem [{\citenamefont {Artemova}\ \emph {et~al.}(1996)\citenamefont
  {Artemova}, \citenamefont {Bj{\"o}rnsson},\ and\ \citenamefont
  {Novikov}}]{artemova}%
  \BibitemOpen
  \bibfield  {author} {\bibinfo {author} {\bibfnamefont {I.~V.}\ \bibnamefont
  {Artemova}}, \bibinfo {author} {\bibfnamefont {G.}~\bibnamefont
  {Bj{\"o}rnsson}}, \ and\ \bibinfo {author} {\bibfnamefont {I.~D.}\
  \bibnamefont {Novikov}},\ }\href@noop {} {\bibfield  {journal} {\bibinfo
  {journal} {ApJ}\ }\textbf {\bibinfo {volume} {461}},\ \bibinfo {pages} {565}
  (\bibinfo {year} {1996})}\BibitemShut {NoStop}%
\bibitem [{\citenamefont {Nowak}\ and\ \citenamefont {Wagoner}(1991)}]{nowak}%
  \BibitemOpen
  \bibfield  {author} {\bibinfo {author} {\bibfnamefont {M.~A.}\ \bibnamefont
  {Nowak}}\ and\ \bibinfo {author} {\bibfnamefont {R.~V.}\ \bibnamefont
  {Wagoner}},\ }\href@noop {} {\bibfield  {journal} {\bibinfo  {journal} {ApJ}\
  }\textbf {\bibinfo {volume} {378}},\ \bibinfo {pages} {656} (\bibinfo {year}
  {1991})}\BibitemShut {NoStop}%
\bibitem [{\citenamefont {Semer{\'a}k}\ and\ \citenamefont
  {Karas}(1999)}]{semerak1999}%
  \BibitemOpen
  \bibfield  {author} {\bibinfo {author} {\bibfnamefont {O.}~\bibnamefont
  {Semer{\'a}k}}\ and\ \bibinfo {author} {\bibfnamefont {V.}~\bibnamefont
  {Karas}},\ }\href@noop {} {\bibfield  {journal} {\bibinfo  {journal} {A\&A}\
  }\textbf {\bibinfo {volume} {343}},\ \bibinfo {pages} {325} (\bibinfo {year}
  {1999})}\BibitemShut {NoStop}%
\bibitem [{\citenamefont {Mukhopadhyay}(2002)}]{mukhopadhyay2002}%
  \BibitemOpen
  \bibfield  {author} {\bibinfo {author} {\bibfnamefont {B.}~\bibnamefont
  {Mukhopadhyay}},\ }\href@noop {} {\bibfield  {journal} {\bibinfo  {journal}
  {ApJ}\ }\textbf {\bibinfo {volume} {581}},\ \bibinfo {pages} {427} (\bibinfo
  {year} {2002})}\BibitemShut {NoStop}%
\bibitem [{\citenamefont {Mukhopadhyay}\ and\ \citenamefont
  {Misra}(2003)}]{mukhopadhyay2003}%
  \BibitemOpen
  \bibfield  {author} {\bibinfo {author} {\bibfnamefont {B.}~\bibnamefont
  {Mukhopadhyay}}\ and\ \bibinfo {author} {\bibfnamefont {R.}~\bibnamefont
  {Misra}},\ }\href@noop {} {\bibfield  {journal} {\bibinfo  {journal} {ApJ}\
  }\textbf {\bibinfo {volume} {582}},\ \bibinfo {pages} {347} (\bibinfo {year}
  {2003})}\BibitemShut {NoStop}%
\bibitem [{\citenamefont {Chakrabarti}\ and\ \citenamefont
  {Mondal}(2006)}]{chakrabarti2006}%
  \BibitemOpen
  \bibfield  {author} {\bibinfo {author} {\bibfnamefont {S.~K.}\ \bibnamefont
  {Chakrabarti}}\ and\ \bibinfo {author} {\bibfnamefont {S.}~\bibnamefont
  {Mondal}},\ }\href@noop {} {\bibfield  {journal} {\bibinfo  {journal}
  {MNRAS}\ }\textbf {\bibinfo {volume} {369}},\ \bibinfo {pages} {976}
  (\bibinfo {year} {2006})}\BibitemShut {NoStop}%
\bibitem [{\citenamefont {Ghosh}\ and\ \citenamefont
  {Mukhopadhyay}(2007)}]{ghosh2007}%
  \BibitemOpen
  \bibfield  {author} {\bibinfo {author} {\bibfnamefont {S.}~\bibnamefont
  {Ghosh}}\ and\ \bibinfo {author} {\bibfnamefont {B.}~\bibnamefont
  {Mukhopadhyay}},\ }\href@noop {} {\bibfield  {journal} {\bibinfo  {journal}
  {ApJ}\ }\textbf {\bibinfo {volume} {667}},\ \bibinfo {pages} {367} (\bibinfo
  {year} {2007})}\BibitemShut {NoStop}%
\bibitem [{\citenamefont {Wegg}(2012)}]{wegg}%
  \BibitemOpen
  \bibfield  {author} {\bibinfo {author} {\bibfnamefont {C.}~\bibnamefont
  {Wegg}},\ }\href@noop {} {\bibfield  {journal} {\bibinfo  {journal} {ApJ}\
  }\textbf {\bibinfo {volume} {749}},\ \bibinfo {pages} {183} (\bibinfo {year}
  {2012})}\BibitemShut {NoStop}%
\bibitem [{\citenamefont {Sarkar}\ \emph {et~al.}(2014)\citenamefont {Sarkar},
  \citenamefont {Ghosh},\ and\ \citenamefont {Bhadra}}]{sarkar}%
  \BibitemOpen
  \bibfield  {author} {\bibinfo {author} {\bibfnamefont {T.}~\bibnamefont
  {Sarkar}}, \bibinfo {author} {\bibfnamefont {S.}~\bibnamefont {Ghosh}}, \
  and\ \bibinfo {author} {\bibfnamefont {A.}~\bibnamefont {Bhadra}},\
  }\href@noop {} {\bibfield  {journal} {\bibinfo  {journal} {Phys. Rev. D}\
  }\textbf {\bibinfo {volume} {90}},\ \bibinfo {pages} {063008} (\bibinfo
  {year} {2014})}\BibitemShut {NoStop}%
\bibitem [{\citenamefont {Tejeda}\ and\ \citenamefont
  {Rosswog}(2014)}]{tejeda2}%
  \BibitemOpen
  \bibfield  {author} {\bibinfo {author} {\bibfnamefont {E.}~\bibnamefont
  {Tejeda}}\ and\ \bibinfo {author} {\bibfnamefont {S.}~\bibnamefont
  {Rosswog}},\ }\href@noop {} {\bibfield  {journal} {\bibinfo  {journal}
  {arXiv:1402.1171}\ } (\bibinfo {year} {2014})}\BibitemShut {NoStop}%
\bibitem [{\citenamefont {Ghosh}\ \emph {et~al.}(2014)\citenamefont {Ghosh},
  \citenamefont {Sarkar},\ and\ \citenamefont {Bhadra}}]{ghosh}%
  \BibitemOpen
  \bibfield  {author} {\bibinfo {author} {\bibfnamefont {S.}~\bibnamefont
  {Ghosh}}, \bibinfo {author} {\bibfnamefont {T.}~\bibnamefont {Sarkar}}, \
  and\ \bibinfo {author} {\bibfnamefont {A.}~\bibnamefont {Bhadra}},\
  }\href@noop {} {\bibfield  {journal} {\bibinfo  {journal} {MNRAS}\ }\textbf
  {\bibinfo {volume} {445}},\ \bibinfo {pages} {4463} (\bibinfo {year}
  {2014})}\BibitemShut {NoStop}%
\bibitem [{\citenamefont {Witzany}\ \emph {et~al.}(2015)\citenamefont
  {Witzany}, \citenamefont {Semer{\'a}k},\ and\ \citenamefont
  {Sukov{\'a}}}]{paperIV}%
  \BibitemOpen
  \bibfield  {author} {\bibinfo {author} {\bibfnamefont {V.}~\bibnamefont
  {Witzany}}, \bibinfo {author} {\bibfnamefont {O.}~\bibnamefont
  {Semer{\'a}k}}, \ and\ \bibinfo {author} {\bibfnamefont {P.}~\bibnamefont
  {Sukov{\'a}}},\ }\href@noop {} {\bibfield  {journal} {\bibinfo  {journal}
  {MNRAS}\ }\textbf {\bibinfo {volume} {451}},\ \bibinfo {pages} {1770}
  (\bibinfo {year} {2015})}\BibitemShut {NoStop}%
\bibitem [{\citenamefont {Bonnerot}\ \emph {et~al.}(2016)\citenamefont
  {Bonnerot}, \citenamefont {Rossi}, \citenamefont {Lodato},\ and\
  \citenamefont {Price}}]{bonnerot}%
  \BibitemOpen
  \bibfield  {author} {\bibinfo {author} {\bibfnamefont {C.}~\bibnamefont
  {Bonnerot}}, \bibinfo {author} {\bibfnamefont {E.~M.}\ \bibnamefont {Rossi}},
  \bibinfo {author} {\bibfnamefont {G.}~\bibnamefont {Lodato}}, \ and\ \bibinfo
  {author} {\bibfnamefont {D.~J.}\ \bibnamefont {Price}},\ }\href@noop {}
  {\bibfield  {journal} {\bibinfo  {journal} {MNRAS}\ }\textbf {\bibinfo
  {volume} {455}},\ \bibinfo {pages} {2253} (\bibinfo {year}
  {2016})}\BibitemShut {NoStop}%
\bibitem [{\citenamefont {Guckenheimer}\ and\ \citenamefont
  {Holmes}(1983)}]{guckenheimer}%
  \BibitemOpen
  \bibfield  {author} {\bibinfo {author} {\bibfnamefont {J.}~\bibnamefont
  {Guckenheimer}}\ and\ \bibinfo {author} {\bibfnamefont {P.}~\bibnamefont
  {Holmes}},\ }\href@noop {} {\emph {\bibinfo {title} {Nonlinear oscillations,
  dynamical systems, and bifurcations of vector fields}}}\ (\bibinfo
  {publisher} {Springer Science \& Business Media},\ \bibinfo {year}
  {1983})\BibitemShut {NoStop}%
\bibitem [{\citenamefont {Abramowicz}\ \emph {et~al.}(1988)\citenamefont
  {Abramowicz}, \citenamefont {Carter},\ and\ \citenamefont
  {Lasota}}]{abralascart}%
  \BibitemOpen
  \bibfield  {author} {\bibinfo {author} {\bibfnamefont {M.~A.}\ \bibnamefont
  {Abramowicz}}, \bibinfo {author} {\bibfnamefont {B.}~\bibnamefont {Carter}},
  \ and\ \bibinfo {author} {\bibfnamefont {J.-P.}\ \bibnamefont {Lasota}},\
  }\href@noop {} {\bibfield  {journal} {\bibinfo  {journal} {Gen. Rel.
  Gravit.}\ }\textbf {\bibinfo {volume} {20}},\ \bibinfo {pages} {1173}
  (\bibinfo {year} {1988})}\BibitemShut {NoStop}%
\bibitem [{\citenamefont {Abramowicz}\ \emph {et~al.}(1997)\citenamefont
  {Abramowicz}, \citenamefont {Lanza}, \citenamefont {Miller},\ and\
  \citenamefont {Sonego}}]{AbraCurv}%
  \BibitemOpen
  \bibfield  {author} {\bibinfo {author} {\bibfnamefont {M.~A.}\ \bibnamefont
  {Abramowicz}}, \bibinfo {author} {\bibfnamefont {A.}~\bibnamefont {Lanza}},
  \bibinfo {author} {\bibfnamefont {J.~C.}\ \bibnamefont {Miller}}, \ and\
  \bibinfo {author} {\bibfnamefont {S.}~\bibnamefont {Sonego}},\ }\href@noop {}
  {\bibfield  {journal} {\bibinfo  {journal} {Gen. Rel. Gravit.}\ }\textbf
  {\bibinfo {volume} {29}},\ \bibinfo {pages} {1585} (\bibinfo {year}
  {1997})}\BibitemShut {NoStop}%
\bibitem [{\citenamefont {Tejeda}\ \emph {et~al.}(2017)\citenamefont {Tejeda},
  \citenamefont {Gafton},\ and\ \citenamefont {Rosswog}}]{tejeda17}%
  \BibitemOpen
  \bibfield  {author} {\bibinfo {author} {\bibfnamefont {E.}~\bibnamefont
  {Tejeda}}, \bibinfo {author} {\bibfnamefont {E.}~\bibnamefont {Gafton}}, \
  and\ \bibinfo {author} {\bibfnamefont {S.}~\bibnamefont {Rosswog}},\
  }\href@noop {} {\bibfield  {journal} {\bibinfo  {journal} {arXiv:1701.00303,
  submitted to MNRAS}\ } (\bibinfo {year} {2017})}\BibitemShut {NoStop}%
\bibitem [{\citenamefont {Laguna}\ \emph {et~al.}(1993)\citenamefont {Laguna},
  \citenamefont {Miller},\ and\ \citenamefont {Zurek}}]{laguna1993smoothed}%
  \BibitemOpen
  \bibfield  {author} {\bibinfo {author} {\bibfnamefont {P.}~\bibnamefont
  {Laguna}}, \bibinfo {author} {\bibfnamefont {W.~A.}\ \bibnamefont {Miller}},
  \ and\ \bibinfo {author} {\bibfnamefont {W.~H.}\ \bibnamefont {Zurek}},\
  }\href@noop {} {\bibfield  {journal} {\bibinfo  {journal} {ApJ}\ }\textbf
  {\bibinfo {volume} {404}},\ \bibinfo {pages} {678} (\bibinfo {year}
  {1993})}\BibitemShut {NoStop}%
\bibitem [{\citenamefont {Abramowicz}\ \emph {et~al.}(1978)\citenamefont
  {Abramowicz}, \citenamefont {Jaroszynski},\ and\ \citenamefont
  {Sikora}}]{donut}%
  \BibitemOpen
  \bibfield  {author} {\bibinfo {author} {\bibfnamefont {M.}~\bibnamefont
  {Abramowicz}}, \bibinfo {author} {\bibfnamefont {M.}~\bibnamefont
  {Jaroszynski}}, \ and\ \bibinfo {author} {\bibfnamefont {M.}~\bibnamefont
  {Sikora}},\ }\href@noop {} {\bibfield  {journal} {\bibinfo  {journal} {A\&A}\
  }\textbf {\bibinfo {volume} {63}},\ \bibinfo {pages} {221} (\bibinfo {year}
  {1978})}\BibitemShut {NoStop}%
\bibitem [{\citenamefont {Misner}\ \emph {et~al.}(1973)\citenamefont {Misner},
  \citenamefont {Thorne},\ and\ \citenamefont {Wheeler}}]{mtw}%
  \BibitemOpen
  \bibfield  {author} {\bibinfo {author} {\bibfnamefont {C.~W.}\ \bibnamefont
  {Misner}}, \bibinfo {author} {\bibfnamefont {K.~S.}\ \bibnamefont {Thorne}},
  \ and\ \bibinfo {author} {\bibfnamefont {J.~A.}\ \bibnamefont {Wheeler}},\
  }\href@noop {} {\emph {\bibinfo {title} {Gravitation}}}\ (\bibinfo
  {publisher} {Macmillan},\ \bibinfo {year} {1973})\BibitemShut {NoStop}%
\bibitem [{\citenamefont {Weinberg}(1972)}]{weinberg}%
  \BibitemOpen
  \bibfield  {author} {\bibinfo {author} {\bibfnamefont {S.}~\bibnamefont
  {Weinberg}},\ }\href@noop {} {\emph {\bibinfo {title} {Gravitation and
  cosmology: principles and applications of the general theory of
  relativity}}},\ Vol.~\bibinfo {volume} {1}\ (\bibinfo  {publisher} {Wiley New
  York},\ \bibinfo {year} {1972})\BibitemShut {NoStop}%
\bibitem [{\citenamefont {Griffiths}\ and\ \citenamefont
  {Podolsk{\'y}}(2009)}]{grifpod}%
  \BibitemOpen
  \bibfield  {author} {\bibinfo {author} {\bibfnamefont {J.~B.}\ \bibnamefont
  {Griffiths}}\ and\ \bibinfo {author} {\bibfnamefont {J.}~\bibnamefont
  {Podolsk{\'y}}},\ }\href@noop {} {\emph {\bibinfo {title} {Exact space-times
  in Einstein's general relativity}}}\ (\bibinfo  {publisher} {Cambridge
  University Press},\ \bibinfo {year} {2009})\BibitemShut {NoStop}%
\bibitem [{\citenamefont {Bardeen}\ \emph {et~al.}(1972)\citenamefont
  {Bardeen}, \citenamefont {Press},\ and\ \citenamefont {Teukolsky}}]{bardeen}%
  \BibitemOpen
  \bibfield  {author} {\bibinfo {author} {\bibfnamefont {J.~M.}\ \bibnamefont
  {Bardeen}}, \bibinfo {author} {\bibfnamefont {W.~H.}\ \bibnamefont {Press}},
  \ and\ \bibinfo {author} {\bibfnamefont {S.~A.}\ \bibnamefont {Teukolsky}},\
  }\href@noop {} {\bibfield  {journal} {\bibinfo  {journal} {The Astrophysical
  Journal}\ }\textbf {\bibinfo {volume} {178}},\ \bibinfo {pages} {347}
  (\bibinfo {year} {1972})}\BibitemShut {NoStop}%
\bibitem [{\citenamefont {Psaltis}(2008)}]{psaltis}%
  \BibitemOpen
  \bibfield  {author} {\bibinfo {author} {\bibfnamefont {D.}~\bibnamefont
  {Psaltis}},\ }\href@noop {} {\bibfield  {journal} {\bibinfo  {journal}
  {Living Rev. Relat.}\ }\textbf {\bibinfo {volume} {11}},\ \bibinfo {pages}
  {1} (\bibinfo {year} {2008})}\BibitemShut {NoStop}%
\bibitem [{\citenamefont {Pettini}(2007)}]{pettini}%
  \BibitemOpen
  \bibfield  {author} {\bibinfo {author} {\bibfnamefont {M.}~\bibnamefont
  {Pettini}},\ }\href@noop {} {\emph {\bibinfo {title} {Geometry and topology
  in Hamiltonian dynamics and statistical mechanics}}},\ Vol.~\bibinfo {volume}
  {33}\ (\bibinfo  {publisher} {Springer Science \& Business Media},\ \bibinfo
  {year} {2007})\BibitemShut {NoStop}%
\bibitem [{\citenamefont {Gibbons}(2015)}]{gibbons}%
  \BibitemOpen
  \bibfield  {author} {\bibinfo {author} {\bibfnamefont {G.}~\bibnamefont
  {Gibbons}},\ }\href@noop {} {\bibfield  {journal} {\bibinfo  {journal}
  {Class. Quantum Grav.}\ }\textbf {\bibinfo {volume} {33}},\ \bibinfo {pages}
  {025004} (\bibinfo {year} {2015})}\BibitemShut {NoStop}%
\bibitem [{\citenamefont {Font}\ \emph {et~al.}(1998)\citenamefont {Font},
  \citenamefont {Ib{\'a}nez},\ and\ \citenamefont
  {Papadopoulos}}]{fonthorizon}%
  \BibitemOpen
  \bibfield  {author} {\bibinfo {author} {\bibfnamefont {J.~A.}\ \bibnamefont
  {Font}}, \bibinfo {author} {\bibfnamefont {J.~M.}\ \bibnamefont
  {Ib{\'a}nez}}, \ and\ \bibinfo {author} {\bibfnamefont {P.}~\bibnamefont
  {Papadopoulos}},\ }\href@noop {} {\bibfield  {journal} {\bibinfo  {journal}
  {ApJ Letters}\ }\textbf {\bibinfo {volume} {507}},\ \bibinfo {pages} {L67}
  (\bibinfo {year} {1998})}\BibitemShut {NoStop}%
\end{thebibliography}%


\end{document}